\pdfoutput=1
\documentclass{article}
\usepackage[utf8]{inputenc}
\usepackage[left=2cm, right=2cm, top=2.54cm, bottom=2.54cm]{geometry}
\usepackage{listings}
\usepackage{color}
\usepackage{graphicx}
\usepackage{mathtools}
\usepackage{amsmath}
\usepackage{amssymb}
\usepackage{gensymb}
\usepackage{amsfonts}
\usepackage{physics,esdiff}
\usepackage{dirtytalk}
\usepackage{hyperref}
\usepackage{url}
\usepackage{alltt}
\usepackage[dvipsnames]{xcolor}
\usepackage{adjustbox}
\usepackage{scrextend}
\usepackage{subfig}
\usepackage{float}
\usepackage{cite}
\colorlet{LightRubineRed}{RubineRed!70!}
\colorlet{Mycolor1}{green!10!orange!90!}
\definecolor{Mycolor2}{HTML}{00F9DE}
\definecolor{mypink2}{RGB}{200, 48, 122}
\usepackage{enumitem}
\newlist{MyIndentedList}{itemize}{4}
\setlist[MyIndentedList,1]{%
	label={},
	noitemsep,
	leftmargin=0pt,
}
\setlist[MyIndentedList]{%
	label={},
	noitemsep,
}
\hypersetup{
	colorlinks=true,
	linkcolor=blue,
	filecolor=magenta,      
	urlcolor=MidnightBlue,
    citecolor=red
}

\usepackage{fancyhdr}
 
\fancypagestyle{firstpage}{%
  \lhead{A Preprint}
  \rhead{Journal of Consideration}
}

\renewcommand{\baselinestretch}{1.5} 

\usepackage[symbol]{footmisc}

\begin{document}
\thispagestyle{firstpage}
\renewcommand{\baselinestretch}{2}
\noindent\textbf{\huge Parametric Analysis of Smartphone Camera for a Low Cost Particle Image Velocimetry\vspace{0.5mm}\\ System}\\
\renewcommand{\baselinestretch}{1.2}
\begin{addmargin}[6em]{5em}
{\bf Vishesh Kashyap$^{1}$\footnote[2]{Vishesh Kashyap and Sushrut Kumar have the Co-First/Equal
authorship.}, Sushrut Kumar$^{1}$, Nehal Amit Jajal$^{2}$, Mrudang Mathur$^{3}$ \& Raj Kumar Singh$^{1}$\footnote[3]{Corresponding Author E-mail : rajkumarsingh@dce.ac.in}}\\
{$^{1}$Department of Mechanical, Production \& Industrial and Automobile Engineering, Delhi Technological University, Delhi, India\\$^{2}$Department of Mechanical and Aerospace Engineering, Ohio State University, U.S.A.\\$^{3}$Department of Mechanical Engineering, University of Texas at Austin, U.S.A. }\\\\
\textbf{Abstract.} This study focuses on assessing smartphone camera characteristics for developing an economic smartphone-based Particle Image Velocimetry (PIV) system. In the investigation, flow around a cylinder was visualized using two commercially-available smartphones (OnePlus 5T and iPhone X) cameras and low-intensity laser diodes. Hydrogen bubbles generated from electrolysis (termed Bubble Image Velocimetry) of aluminum electrodes were used as seeding medium. OpenPIV, an open-source toolbox, was used for processing captured images and obtaining the flow fields. A parametric analysis of the two smartphones was conducted across varying camera characteristics such as ISO, exposure compensation and frame rate. The results obtained through experimentation were compared with the results of a validated computational fluid dynamics (CFD) study with the same flow conditions and were found to be in good agreement, with deviation ranging from 1\% to 3.5\% for iPhone X and 1\% to 7\% for OnePlus 5T. It was observed that a higher frame rate results in greater accuracy of the measurement. Further, an exposure compensation of -1 EV and an ISO of 400 was found to produce results with the least error as compared to CFD values.\\
\end{addmargin}
\section{Introduction}

\begin{figure}[H]
\centering
\includegraphics[width=1\textwidth]{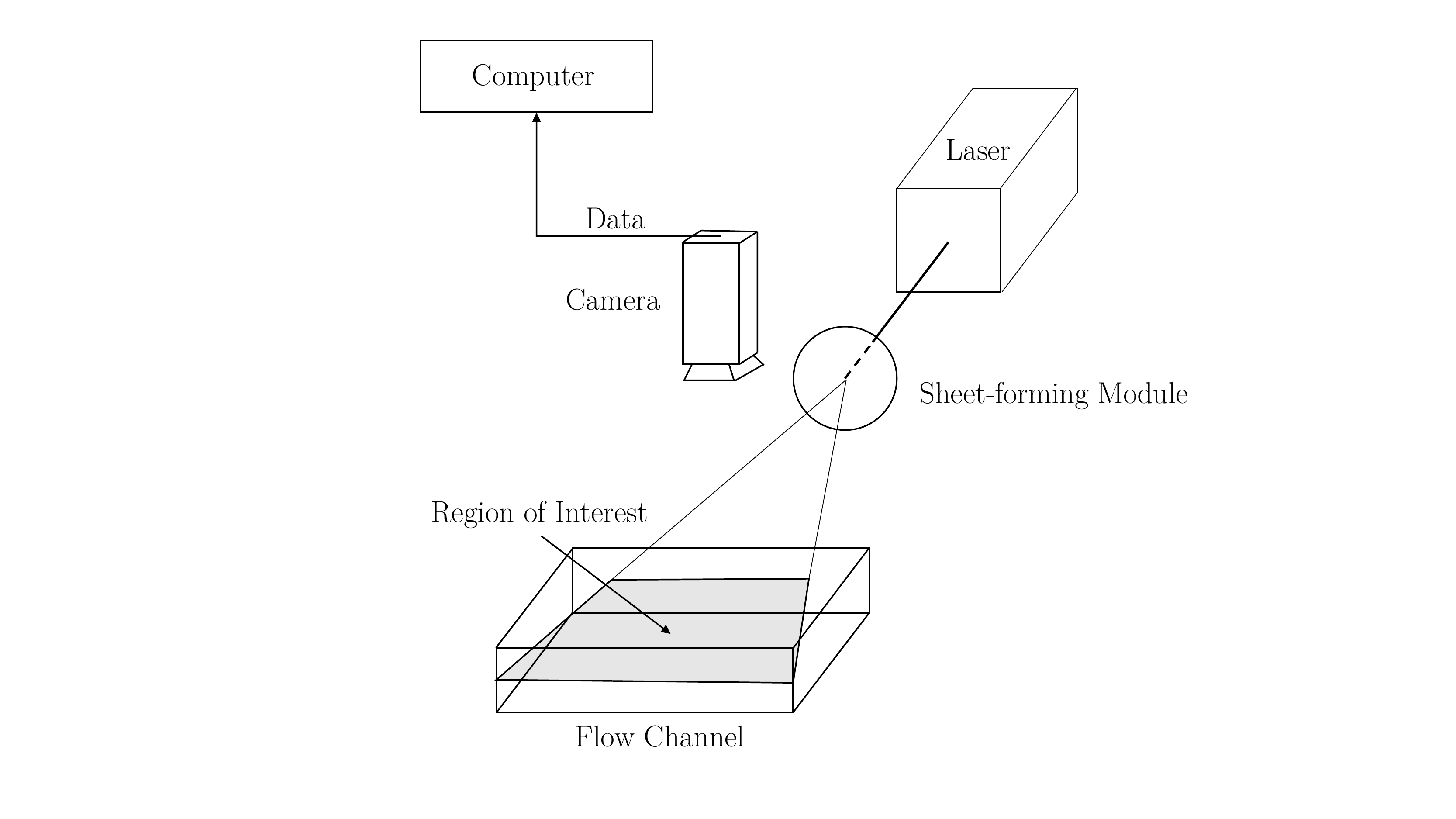}
\caption{Basic components of a PIV system}
\label{f1}
\end{figure}

Experimental fluid mechanics routinely requires the quantitative visualization and measurement of the overall flow properties of a fluid domain \cite{s1}. However, while flow measurement techniques such as Hot-Wire and Laser Doppler Anemometry are regularly used for measurements of such properties, they are restricted to measuring only at a few points of the domain simultaneously.\\
Particle Image Velocimetry (PIV) is an experimental technique that involves the visualization of fluid flow under given experimental conditions. The main components of a PIV system are a laser, a camera, seeding particles and the channel, as shown in Figure \ref{f1}. The fluid flowing through the channel is seeded with neutrally buoyant particles that remain suspended within it. These particles are illuminated by a plane of light originating from the laser and converted into a sheet by passing through an arrangement of cylindrical lenses. The flow of the seeding particles on the illuminated plane is analyzed using a correlation-based algorithm which determines the average motion of particles within small regions of the plane known as interrogation windows, between image pairs \cite{s2,s3,s4}. It then provides the flow profiles of the particles in form of vector lines. It may also provide parameters such as vorticity, given appropriate inputs such as time difference and reference distance \cite{s5}.\\
Measurements   using   PIV   must   be   of   high accuracy and resolution for them to be usable for the calculation of quantities such as velocity, vorticity and other secondary variables \cite{s6,s7,s8,s9,s10}. However, one  major  hindrance to the application for PIV in research  has  been  the  cost  of the system. Research grade systems often cost upwards of \$100,000. This makes them inaccessible to a great number of researchers and students \cite{s11}.\\

A number of considerations are integral to the reduction in cost of a PIV system. The seeding particles resist flow, and thus must be an appropriate size and density. These characteristics can be based on the resolution of the camera \cite{s12}. To ensure accurate measurement, a greater spatial concentration of seeding particles is required in the region of measurement. Materials such as $TiO_{2}$ \cite{s13}, $Al_{2}O_{3}$ \cite{s14}, Polystyrene  \cite{s15}, resin \cite{s16} and microspheres \cite{s17} have in the past been used as seeding particles, primarily due to their non-corrosive, non-abrasive and chemically inert nature, with diameters of upto 0.5 mm \cite{s18,s19}. Dieter et al. \cite{s20} utilized hydrogen bubbles obtained by electrolysis of water in order to produce Eulerian flow vectors close to the air-water interface. Expensive seed particles have also attempted to be replaced by less costly alternatives such as fluorescent particles mixed with fluorescent dye \cite{s21} and white polystyrene beads \cite{s22}. In many of these solutions, the availability of the alternative is location-specific and sparse.\\

Commercial PIV applications typically utilize high-powered Nd:YAG (neodymium-doped yttrium aluminum garnet) \cite{s23,s24}, ion \cite{s25} or argon \cite{s26,s27} lasers to illuminate the area of measurement. The pulse generation of these lasers is typically synchronized with the camera frame rate using separate equipment known as synchronizers. However, the price of these lasers is very high \cite{s28}. High-powered LEDs \cite{s29,s30,s31,s32} with outputs over 10 W have been used to provide illumation as a cheaper option. Ring and Lemley \cite{s33} demonstrated the use of inexpensive laser diodes in PIV applications.\\

Typical modern PIV cameras contain a CCD or CMOS chip, which allows for the capturing of high-quality images and the reduction of image noise. Depending on the type of analysis, the frame rates of such cameras can be as high as 25600 frames-per-second (FPS). Modern PIV systems contain cameras with additional features such as multiple exposures, small pixel sizes (px) and short interframe time. However, these cameras are very expensive with costs upwards of \$10000 common in the market. In recent years, similar features have become common in smartphone cameras, though at a much lower level. The OnePlus 5T (M1) and the iPhone X (M2) represented the highest end of camera features at the time experimentation was carried out. M2 has two 10 megapixel (MP) cameras, one each with a pixel size of 1.22 micron and 1.0 micron. M1 has pixel sizes of 1.12 micron and 1.0 micron. At a resolution of 1920 x 1080 px, M2 has a maximum frame rate of 240 FPS while M1 has a maximum of 60 FPS.\\

Adjusting the ISO or light sensitivity controls the amount of light the camera gets in order to form the image. A high ISO can introduce into the image noise in the form of extra grain. The PIV apparatus is operated in a darkened room, and hence the camera automatically sets ISO to high values. The automatic ISO value for the current study was observed to be in excess of 1000, which led to very bright images.   The frame rate of the video defines the number of frames the camera captures per second. For the purpose of PIV measurements, a high frame rate is considered better since the relative movement of the seed particles among consecutive frames can be perceived better. Exposure compensation allows the user to override the common exposure settings chosen by the camera in favor of brighter or darker images. PIV typically requires two exposures in order that each may be isolated and cross-correlation analysis may be performed \cite{s42}. In smartphone cameras, adjustment of exposure is done by adjusting the exposure time, i.e. the time interval at which the two exposure shots are taken. While initial smartphone cameras did not allow for this, modern cameras can be used to adjust exposure compensation, which may increase or decrease the exposure time.\\

Experimentation in an open channel allows for greater ease in the replacement of test bodies around which flow of liquid is to be visualized. A number of experimental studies of the velocity profile of developed open channel flow have been made \cite{s34,s35}. Various studies have been conducted over the decades to calculate the establishment length for an open channel under different conditions, yet there is limited information available to reach a parametric relation for the same \cite{s35}.\\

While previous studies have attempted to develop economical alternatives to various elements of a PIV system \cite{s5,s11,s21,s28} minimal research is available on the minimization of cost of the system along with an analysis of recording alternatives. This study involves the design and development of an inexpensive, yet accurate smartphone-based PIV system to be used for educational and research purposes. Hydrogen bubbles produced from the electrolysis of aluminum electrodes were used as seeding particles, commonly available laser diodes of power 1.5 Watt (W) for illumination and a smartphone camera for recording. The open-source PIV software OpenPIV \cite{s37} was used for analysis. The system has the advantage of convenient availability of components, ease of operation, repeatability and high adaptability. Measurements were carried out with a smartphone camera, at varying properties such as ISO, exposure compensation and frame rate.

\section{Methodology}
\subsection{Design and Fabrication of Apparatus}
The apparatus was designed and fabricated as well as experiments conducted at the Fluid Systems Laboratory, Delhi Technological University. The primary components of the apparatus were the channel for flow of water, electrodes for the production of hydrogen bubbles, pump for circulation and reservoirs for the provision of appropriate head and water collection.\\

An analytical study was carried out to determine the dimensions of the channel. Durst et al. \cite{s38} developed a parametric relation between the length of the channel (L) and the free surface level (D) in terms of the Reynolds Number (Re), as per equation \eqref{eq1}.\\
\begin{align}
\frac{L}{D}=[0.631^{1.6}+(0.0442Re)^{1.6}]^{\frac{1}{1.6}}\label{eq1}
\end{align}\\

\begin{figure}[H]
\centering
\includegraphics[width=1\textwidth]{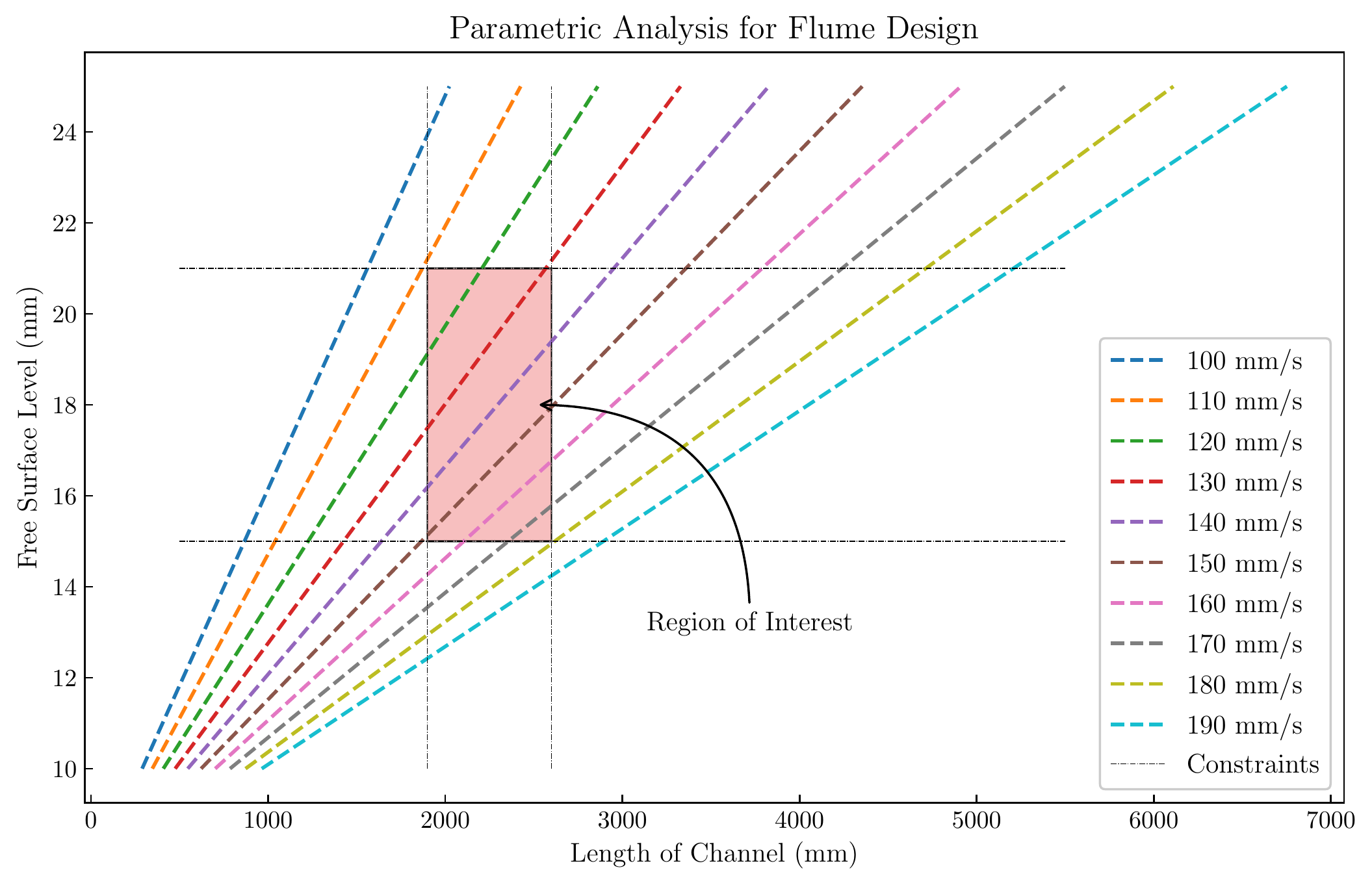}
\caption{Analytical study for flume design}
\label{f2}
\end{figure}

A region of interest was defined by studying velocities between 100 mm/s and 250 mm/s, constraining the length of the channel between 1900 and 2600 mm as per the space available. Figure \ref{f2} shows the development length of the channel plotted against the free surface level at various velocities.\\

\begin{figure}[H]
\centering
\includegraphics[width=1\textwidth]{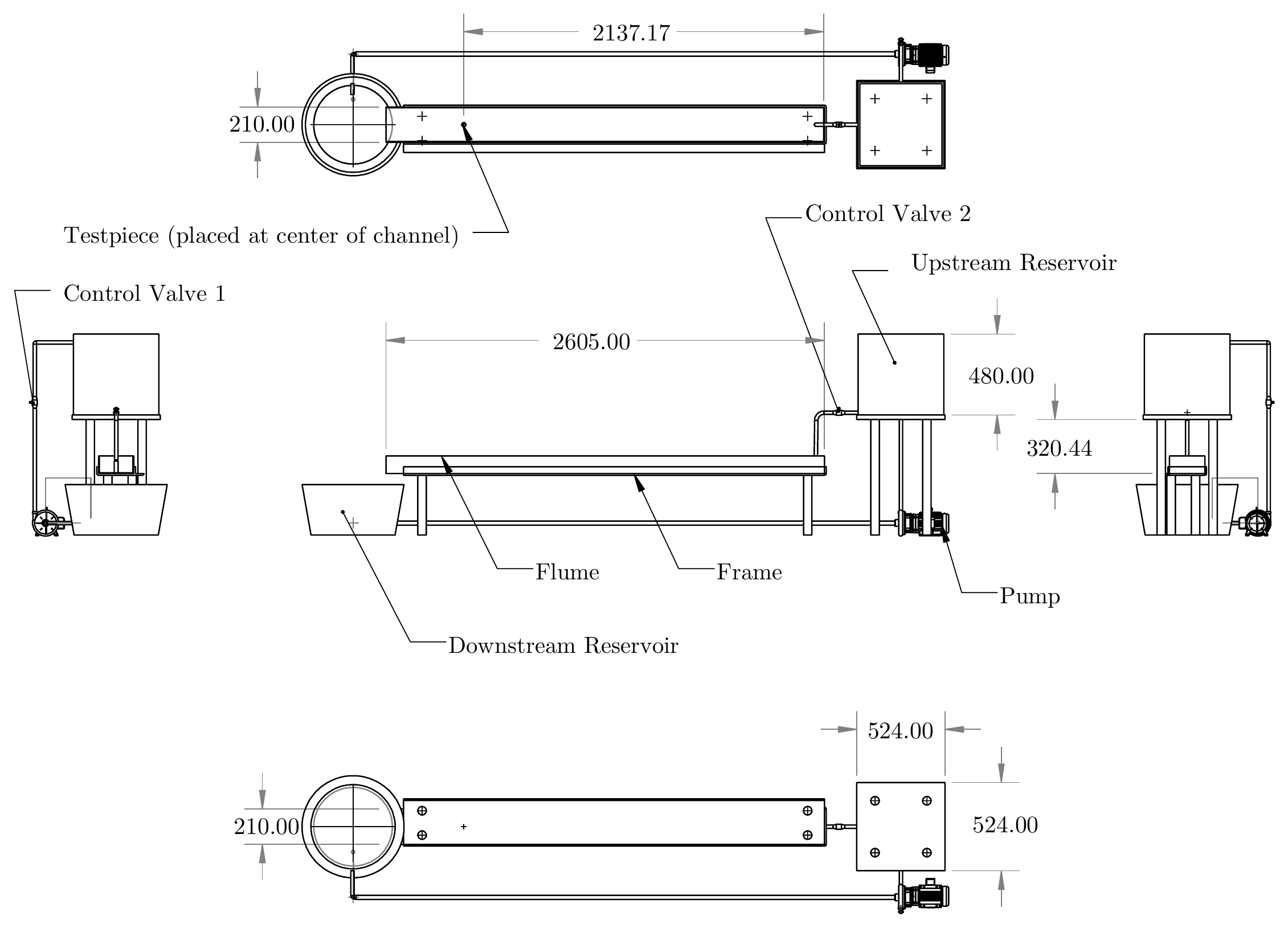}
\caption{Schematic of Developed Apparatus}
\label{f3}
\end{figure}

Based on the constraints, a velocity of 150 mm/s was chosen for further analysis, and a development length of 2137 mm was determined from the graph shown in Figure \ref{f2}. The head of the upstream reservoir was fixed at 350 mm in order to obtain the required velocity of flow. Figure \ref{f3} represents the dimensions of the apparatus. 
The acrylic water channel was constructed through laser-cutting. The length of the water channel was 2.3 m, the width was 0.2 m and the height was 0.1 m. The recirculation was driven by a single-phase, 0.5 HP pump from the reservoir downstream to the one upstream, as shown in figure \ref{f3}.
Two aluminium electrodes of length 200 mm and radius 2.5 mm were placed on either side of the test body across the width of the channel and connected to a 5 W single phase variable transformer.  The cathode was placed at 70 mm upstream and the anode 50 mm downstream from the test body. 50g salt was added to the upstream reservoir in order to increase the rate of electrolysis \cite{s39}. The electrodes were operated at a potential difference of 40 Volt (V). Hydrogen bubbles were generated by electrolysis. 
A 5 mW red (wavelength 650 nm) laser diode was used to illuminate the flow region. The laser module containing sheet-forming lenses was placed so as to create a laser sheet in a direction perpendicular to the flow of water and parallel to the ground. A mount was designed to ensure the parallelism of the plane of the laser sheet with respect to the free surface and the smartphone camera. The smartphone was placed at 100 mm above the base of the channel and parallel to the ground at the position of the test body. The total cost of the experimental setup is INR 17300, and mentioned component-wise in Appendix I.
\subsection{Experimentation}
\begin{figure}[H]
\centering
\includegraphics[width=0.7\textwidth]{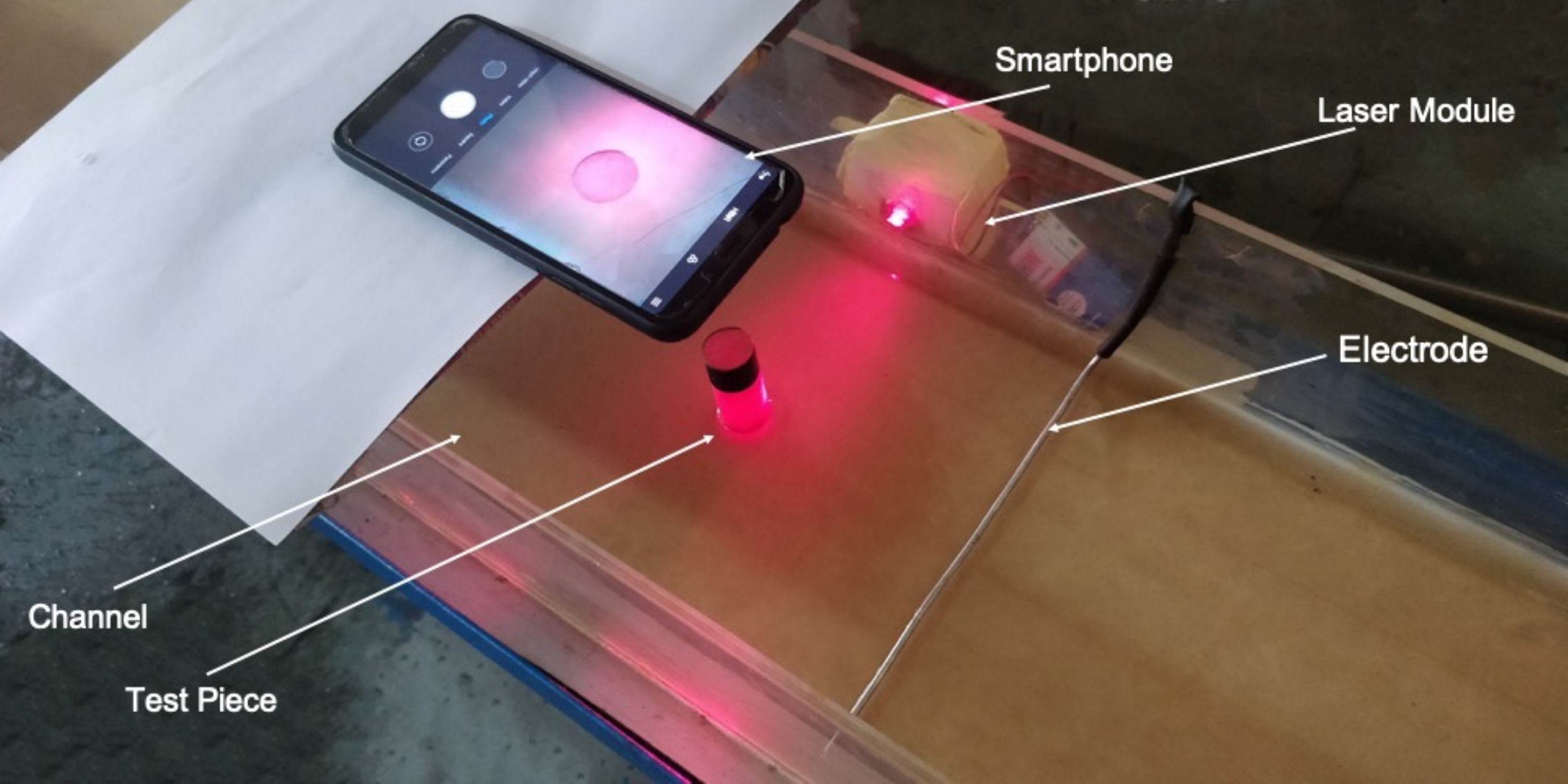}
\caption{Region of experimentation and experimental setup.}
\label{f4}
\end{figure}

The cylindrical test body of radius 20 mm was placed at a distance of 2137 mm downstream. The mount containing the laser was placed to one side of the channel, perpendicular to the direction of flow. The laser sheet was parallel to the surface of water. The camera was placed directly above the specimen, as shown in figure \ref{f4}. The upstream reservoir, downstream reservoir and flume were filled with water up to the required head, and the motor was started. The flow was allowed to stabilize by adjusting the speed of the motor to maintain a constant head in the upstream reservoir. When the flow was steady, a smartphone camera was used to record a video of length 10 seconds (s). The camera’s field of view covered a length of 120 mm along the channel just above the free surface level at a height of 103 mm. At a video resolution of $1920\times1080$ px, this yielded an input of 0.0000625 m/px.\\ 

Videos for parametric analysis of frame rate, exposure compensation and ISO were recorded as per the available settings in the phones, with the following conditions.
\begin{enumerate}
\item \textbf{Frame Rate:} Recordings were made at frame rates of 60 and 30 FPS for M1 and 240 FPS and 60 FPS for M2. The exposure compensation and ISO were kept constant at -1 EV and 400 respectively, at 1080p video mode.
\item \textbf{Exposure Compensation:} Recordings were made at exposure compensations of -2, -1, 0, 1 and 2 EV for M1 and M2. The ISO was kept constant at 400 for both smartphones, while the frame rate was 60 FPS at 1080p video mode.
\item \textbf{ISO:}  Recordings were made at ISO modes of ISO 100, 200, 400 and 800 for M1. The frame rate was kept constant at 60 FPS and the exposure compensation at -1 EV, at 1080p video mode. M2 does not have a setting for variation of ISO.
\end{enumerate}

The conditions are summarized in Table \ref{t1}.
\begin{table}[H]
\centering
\caption{\label{t1}Analysis settings for parametric study}
\begin{tabular}{ |l|l|l|l|l| }
\hline
\textbf{Setting} & \textbf{Phone} & \textbf{Value} & \textbf{Other Settings} & \textbf{No. of Readings}\\
\hline
Frame Rate & M1 & 30, 60 FPS & -1 EV, ISO 400 & 4\\
\hline
Frame Rate & M2 & 60, 240 FPS & -1 EV, ISO 400 & 4\\
\hline
Exposure Compensation & M1 & -2, -1, 0, 1, 2 EV & 60 FPS, ISO 400 & 3\\
\hline
Exposure Compensation & M2 & -2, -1, 0, 1, 2 EV & 60 FPS, ISO 400 & 4\\
\hline
ISO & M1 & ISO 100, 200, 400, 800 & 60 FPS, -1 EV & 5\\
\hline
\end{tabular}
\end{table}
\subsection{Data Acquisition and Analysis}
The software Free Video to JPG Converter \cite{s40} was used to split the recorded video into its constituent frames. The open-source MATLAB\textsuperscript{\textregistered} toolbox OpenPIV was used for the analysis. An interrogation window of size $128\times128$ px was defined based on the number of particles in each window, and a pair of consecutive images was loaded into the toolbox. Metres per pixel (m/px) and time step dt (s) were taken as inputs. The value of velocity is calculated as per equation \eqref{eq2}.\\
\begin{align}
\text{Velocity}(\frac{m}{s}) = \frac{\text{meters per pixel }(\frac{m}{px}) \times \text{number of pixels }(px)}{\text{time step }(s)}\label{eq2}
\end{align}

\begin{figure}[H]
\centering
\includegraphics[width=1\textwidth]{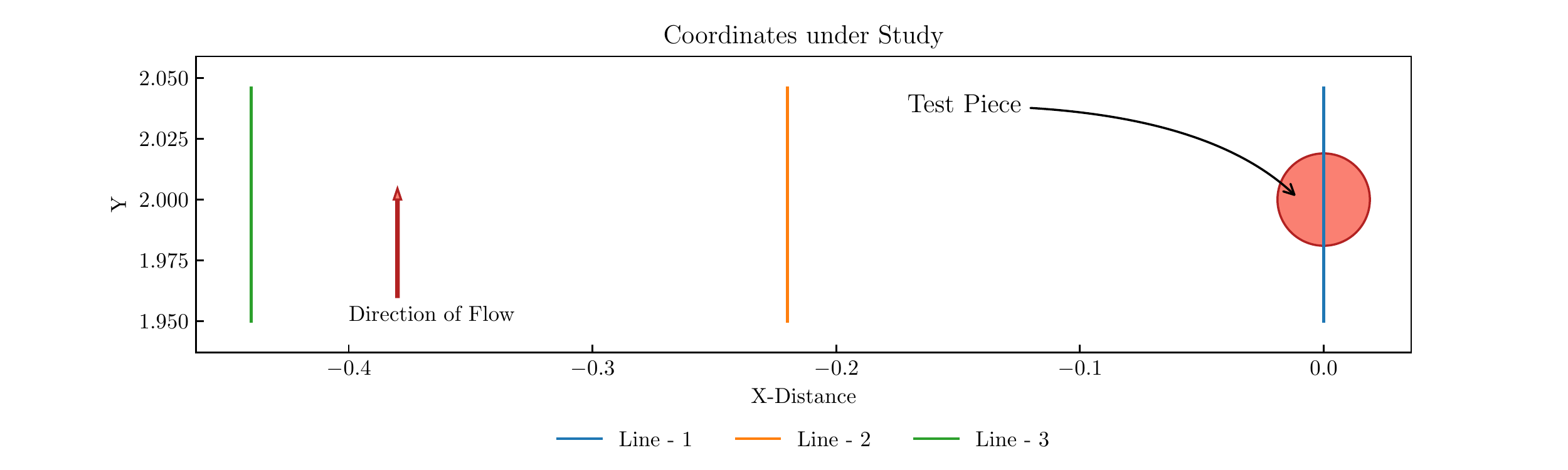}
\caption{Lines studied for analysis.}
\label{f5}
\end{figure}

A vector field is constructed by OpenPIV by comparing the movement of particles between image pairs, and this process is repeated over every interrogation window of the region of interest. The result includes all the vectors normalized in comparison to the free stream velocity. OpenPIV then compiles a data file containing the calculated values of velocity in the X and Y direction of all the points studied.
In this study, velocity values were obtained for a total of 75 points in 3 lines segments along the direction of flow as shown in Figure \ref{f5}. The location of these segments was chosen in order to study the effectiveness of the smartphone cameras to appropriately capture the seeding particles in three distinct regions - at the centerline of the test piece (Line - 1), near the channel wall (Line - 2) and in a region between previous two (Line - 3).
\subsection{Numerical Simulation}
A Computational Fluid Dynamics (CFD) analysis was conducted in order to simulate the flow of water at the conditions of the present study. The analysis was performed by solving the RANS equations \cite{gti}. These are obtained by reynolds averaging the conservation form of transport equation. Equation \ref{cont} shows the continuity equation whereas Equation \ref{momt} is the momentum equation. In the equations, $u_{i}$ is velocity of fluid, $\rho$ is density, $p$ is pressure, $\tau_{ i j }$ represents the Reynolds stress terms and  quantities with overbar are time average means.\\
\begin{align}
\frac { \partial \bar { \rho } } { \partial t } + \frac { \partial \left( \bar { \rho } \hat { u } _ { i } \right) } { \partial x _ { i } } = 0
\label{cont}
\end{align}\\
\begin{align}
\frac { \partial \left( \bar { \rho } \hat { u } _ { i } \right) } { \partial t } + \frac { \partial \left( \bar { \rho } \hat { u } _ { i } \hat { u } _ { j } \right) } { \partial x _ { j } } = - \frac { \partial \bar { p } } { \partial x _ { i } } + \frac { \partial \bar { \sigma } _ { i j } } { \partial x _ { j } } + \frac { \partial \tau _ { i j } } { \partial x _ { j } }
\label{momt}
\end{align}

The transport equations are discretised using the Finite Volume Method. The $k-\omega$ SST turbulence model was used to provide closure for the RANS equations. It is a two equation eddy viscosity model and is shown by Equation \ref{kw1} and \ref{kw2}. The $SST$ (Shear Stress Transport) formulation is helpful in working with high sensitivity of $k-\omega$ with inlet turbulence conditions \cite{gaslat}.\\
\begin{align}
\frac { \partial k } { \partial t } + U _ { j } \frac { \partial k } { \partial x _ { j } } = P _ { k } - \beta ^ { * } k \omega + \frac { \partial } { \partial x _ { j } } \left[ \left( \nu + \sigma _ { k } \nu _ { T } \right) \frac { \partial k } { \partial x _ { j } } \right]
\label{kw1} 
\end{align}\\
\begin{align}
\frac { \partial \omega } { \partial t } + U _ { j } \frac { \partial \omega } { \partial x _ { j } } = \alpha S ^ { 2 } - \beta \omega ^ { 2 } + \frac { \partial } { \partial x _ { j } } \left[ \left( \nu + \sigma _ { \omega } \nu _ { T } \right) \frac { \partial \omega } { \partial x _ { j } } \right] + 2 \left( 1 - F _ { 1 } \right) \sigma _ { \omega 2 } \frac { 1 } { \omega } \frac { \partial k } { \partial x _ { i } } \frac { \partial \omega } { \partial x _ { i } }
\label{kw2}
\end{align}\\

SIMPLE algorithm (Semi-Implicit Method for Pressure Linked Equations) was used for pressure velocity coupling \cite{mru}. The simulations were conducted using ANSYS Fluent Academic Research 16.0. The computational domain follows the exact dimension of the actual apparatus. Grid independence tests were conducted and a cell size of 0.75 mm was determined for analysis. It was also ensured that the value of $\text{y}+$ always remained below $1$ in the vicinity of cylinder and channel wall to be in accordance with law of wall function.\\

\begin{figure}[H]
\centering
\includegraphics[width=0.5\textwidth]{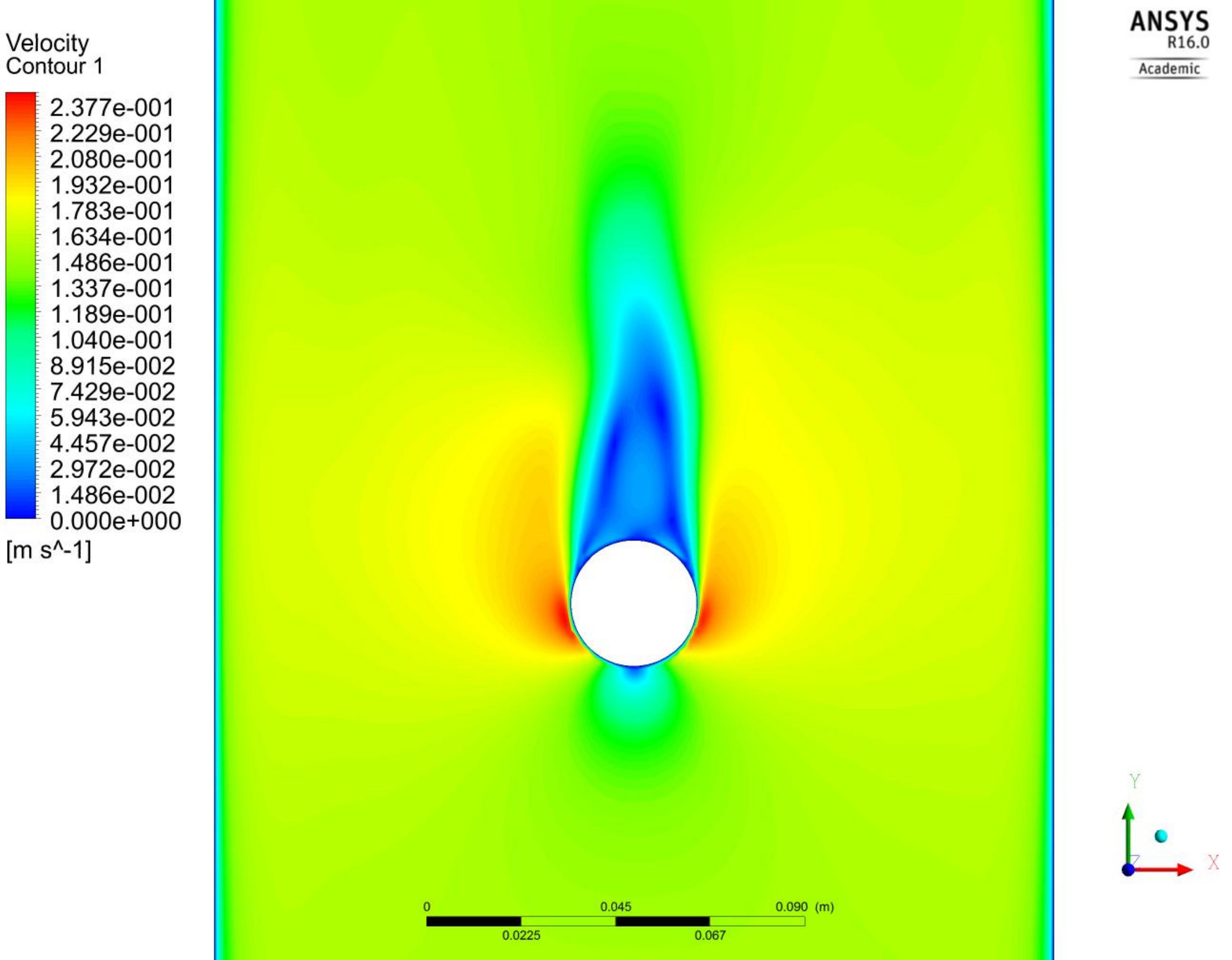}
\caption{Velocity contours from top view of the CFD analysis}
\label{f6}
\end{figure}

The inlet boundary condition was given as a velocity of 0.15 m/s and the outlet as pressure boundary condition with gauge pressure $0$. Moreover, the boundary condition for cylinder and channel wall was set as no-slip. The drag coefficient obtained from the CFD study was observed to be 0.426. For purpose of validating the numerical simulation, this drag coefficient was correlated with previous literature data. As in published literature by Schlichting \cite{s41}  and Pritchard \cite{s42}, where Reynolds Number was plotted against drag coefficient, this value was experimentally found to be 0.420 in the conditions of the present study. These values are found to be within the margin of error ($<1.5\%$). Thus the CFD simulation can be utilised for validating the PIV velocity data obtained from smartphones, M1 \& M2. This is done by determining the velocity data from CFD simulation. The velocity field from the CFD analysis was determined for the three line segments chosen, as shown in Figure \ref{f6}.\\ 

For the purpose of the study, the ordinates ($Y$) of the points under consideration were non dimensionalised as shown in Equation \ref{ystar}. \\
\begin{align}
Y^{*} = \frac{Y}{Y_\text{max}-Y_\text{center}}\label{ystar}
\end{align}\\

\begin{figure}[H]
\centering
\includegraphics[width=0.4\textwidth]{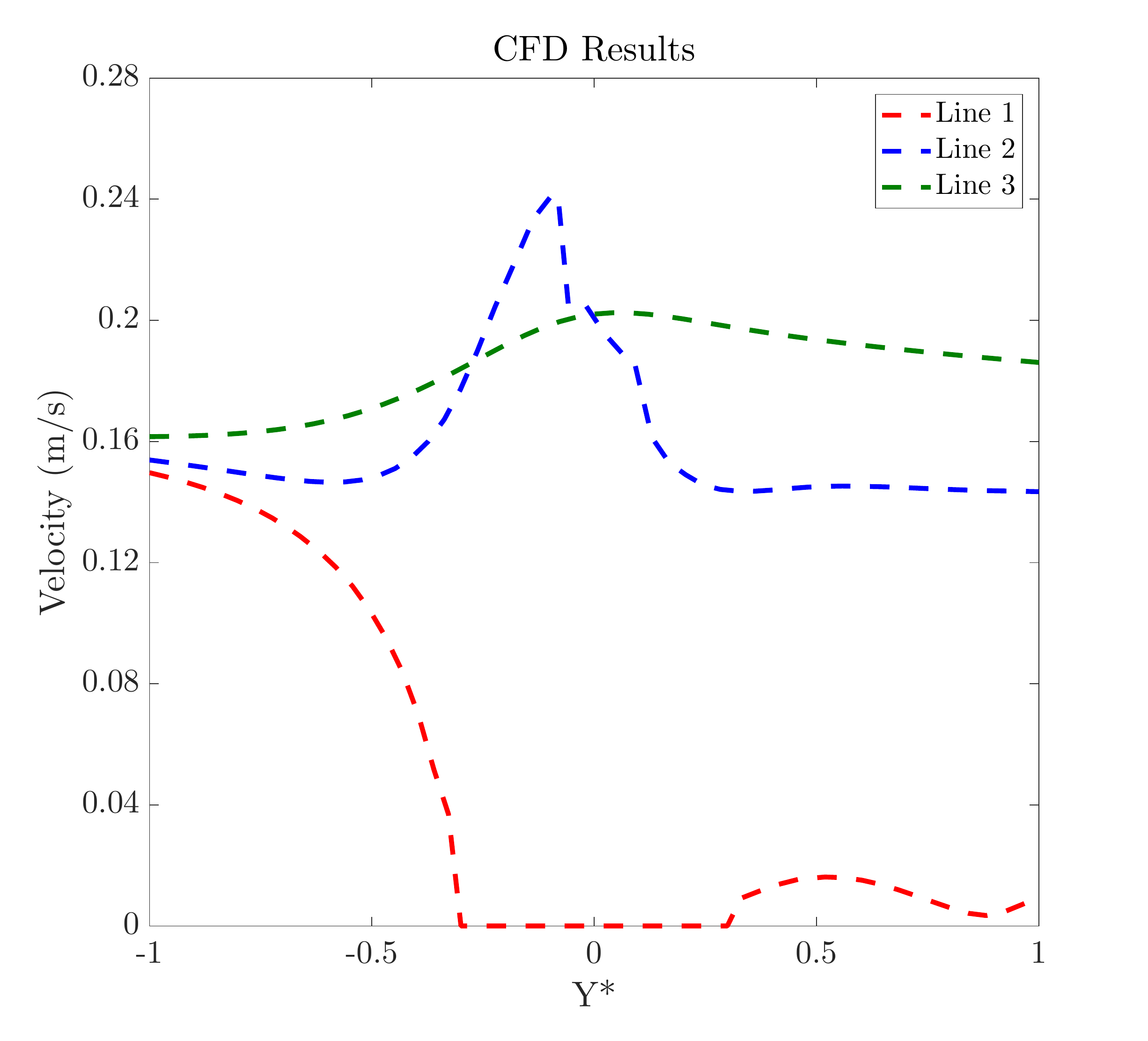}
\caption{Velocity values plotted with $Y^{*}$}
\label{f6b}
\end{figure}

Where Y is the ordinate of the point under consideration, and $Y_\text{max}$, is the maximum variation of the line segment from its central value and $Y_\text{center}$ is the value at the center of the line segment. The velocities were plotted in Figure \ref{f6b} against values of $Y^{*}$.
\section{Results and Discussion}
\subsection{Analysis of Frame Rate}
\begin{figure}[H]
\centering
\begin{minipage}{.5\linewidth}
\centering
\subfloat[]{\label{f7a}\includegraphics[scale=.32]{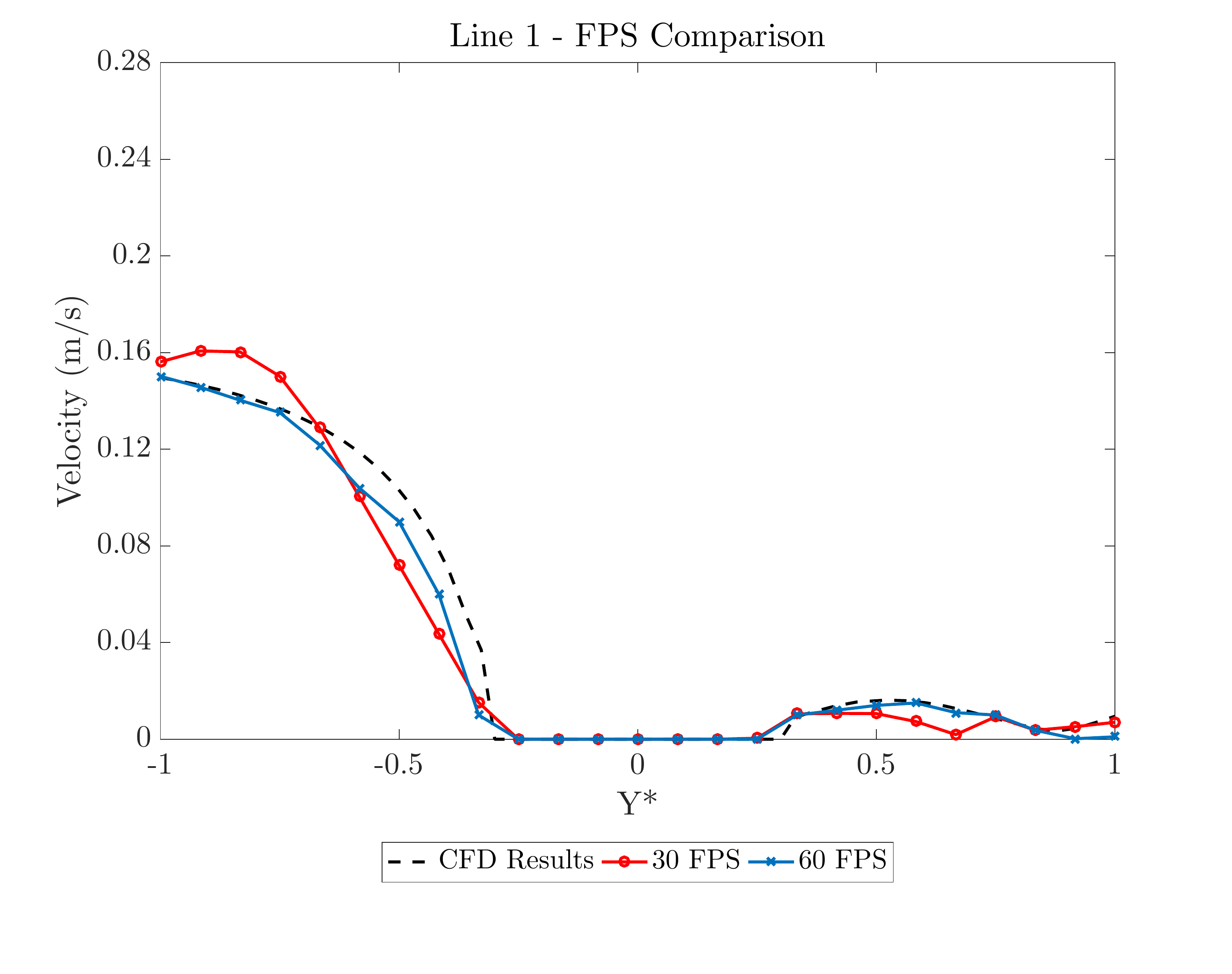}}
\end{minipage}%
\begin{minipage}{.5\linewidth}
\centering
\subfloat[]{\label{f7b}\includegraphics[scale=.32]{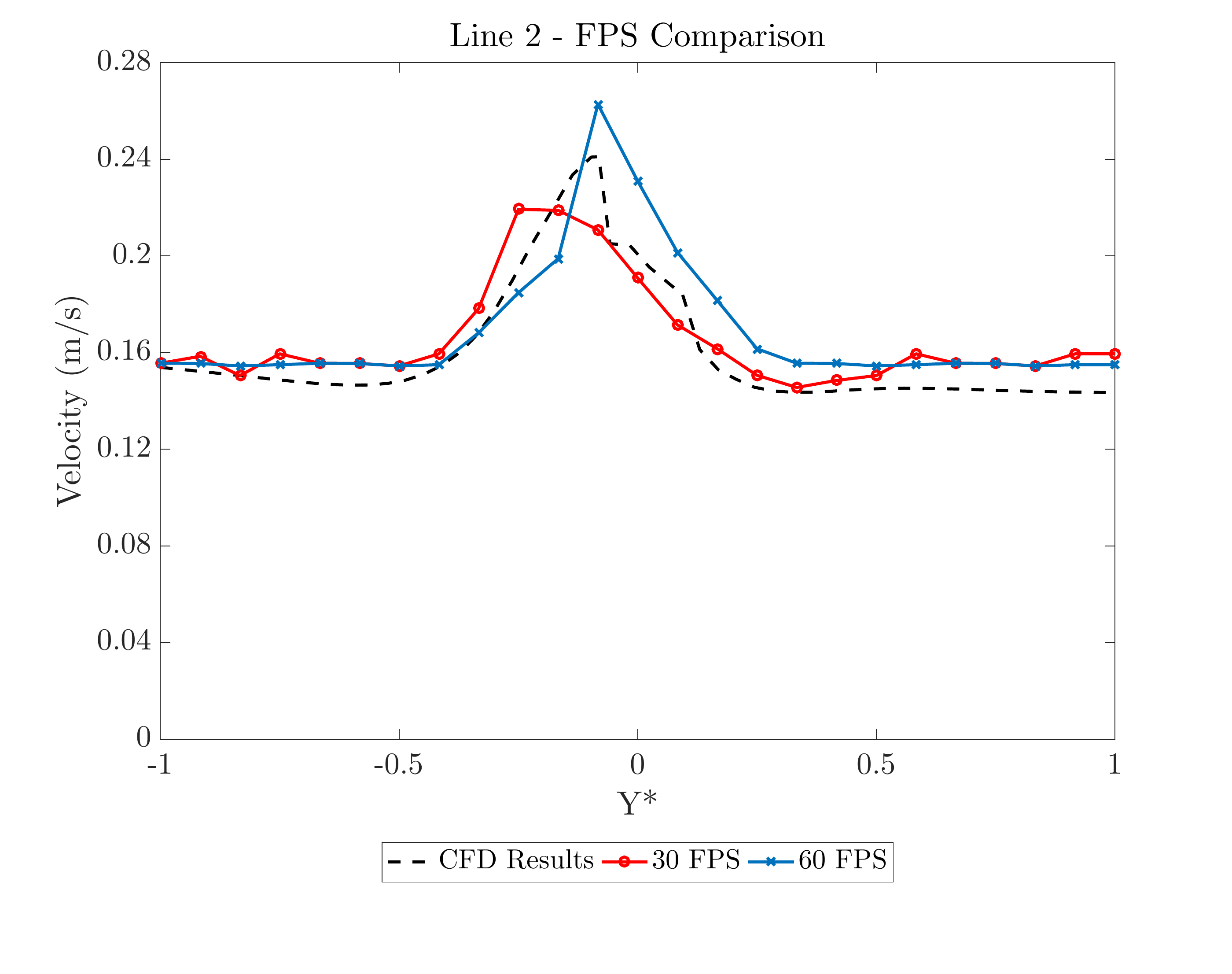}}
\end{minipage}\par\medskip
\centering
\subfloat[]{\label{f7c}\includegraphics[scale=.32]{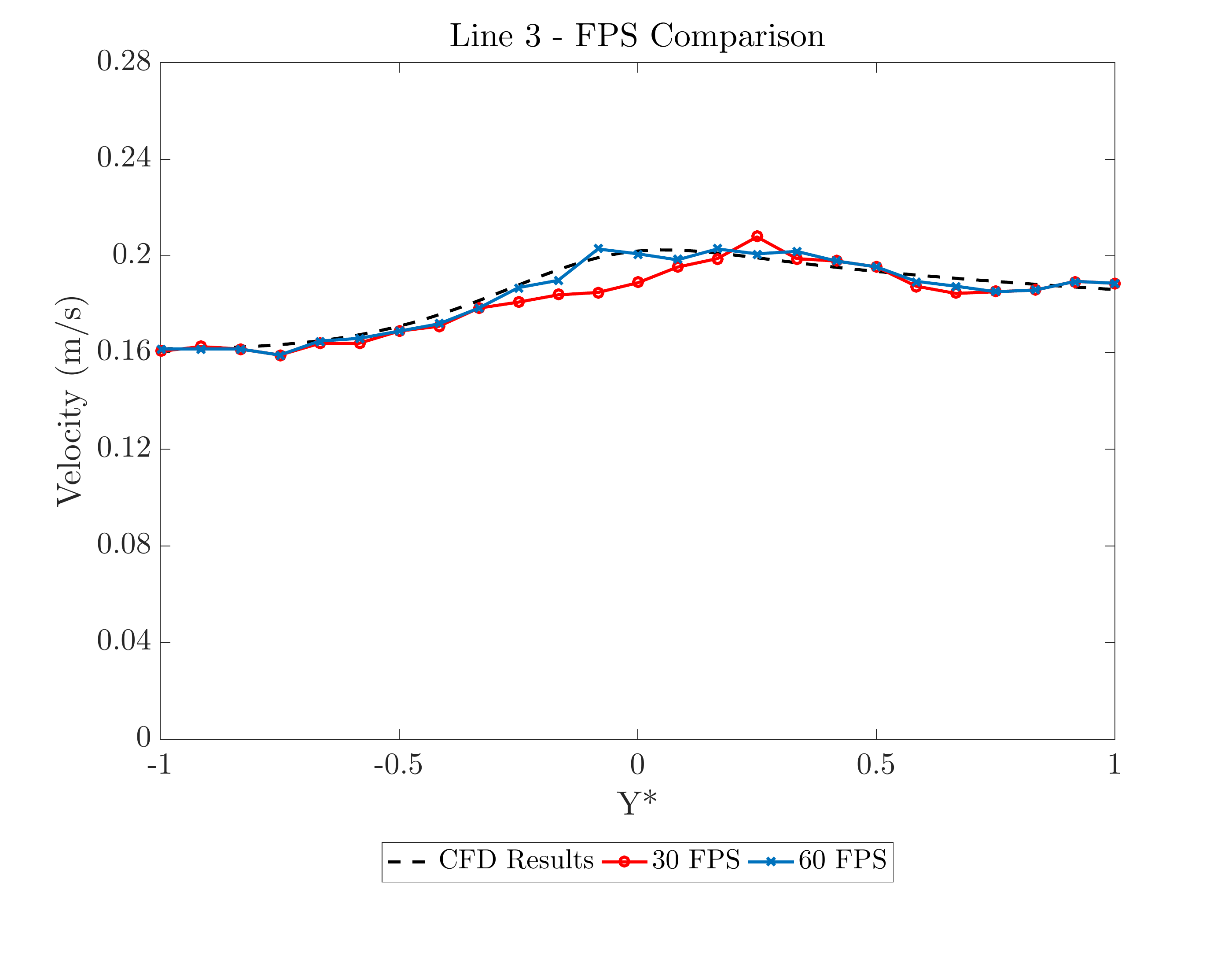}}
\caption{Values of velocities at various frame rates with change in Y-coordinate for M1}
\label{f7}
\end{figure}

\begin{figure}[H]
\centering
\begin{minipage}{.5\linewidth}
\centering
\subfloat[]{\label{f8a}\includegraphics[scale=.32]{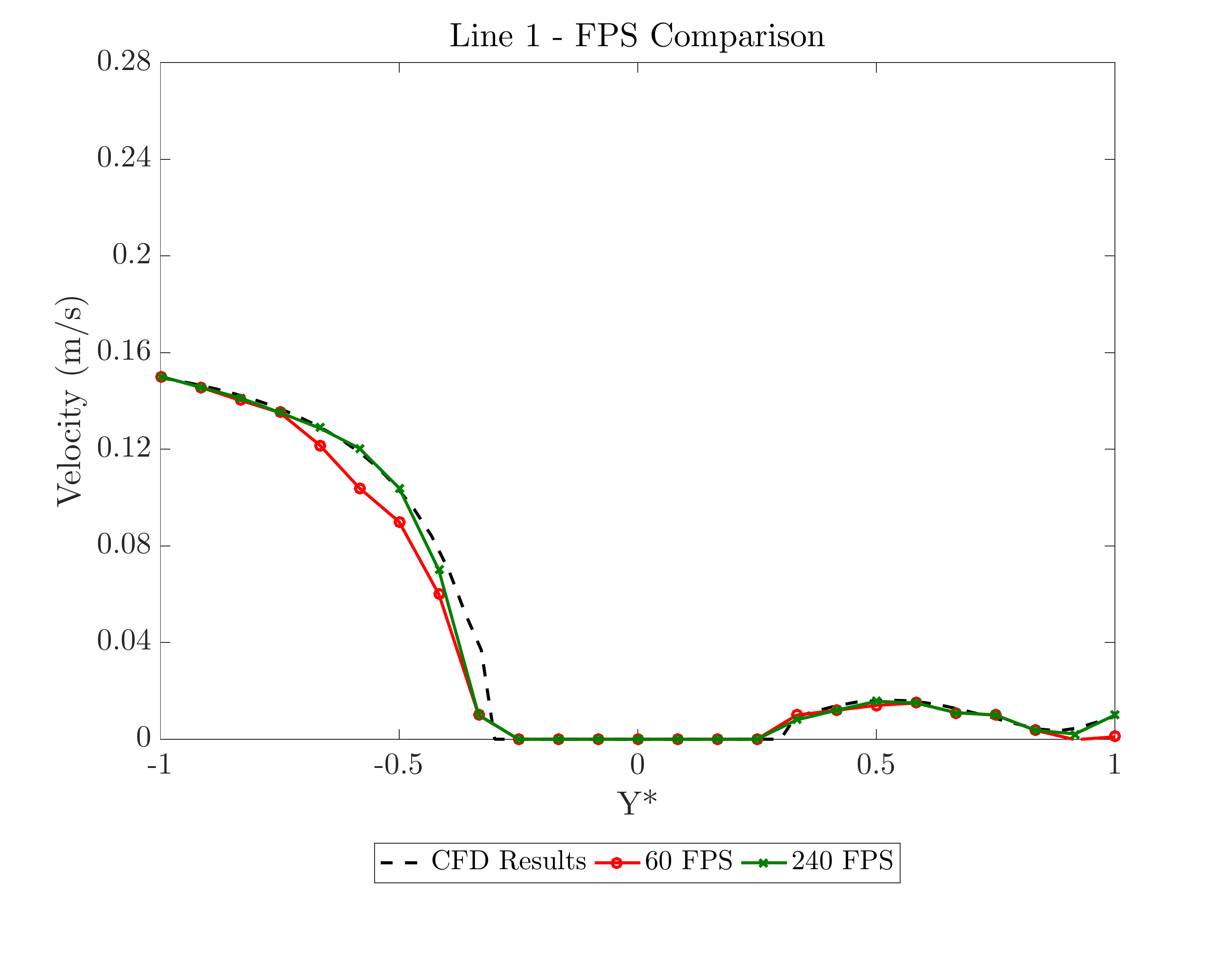}}
\end{minipage}%
\begin{minipage}{.5\linewidth}
\centering
\subfloat[]{\label{f8b}\includegraphics[scale=.32]{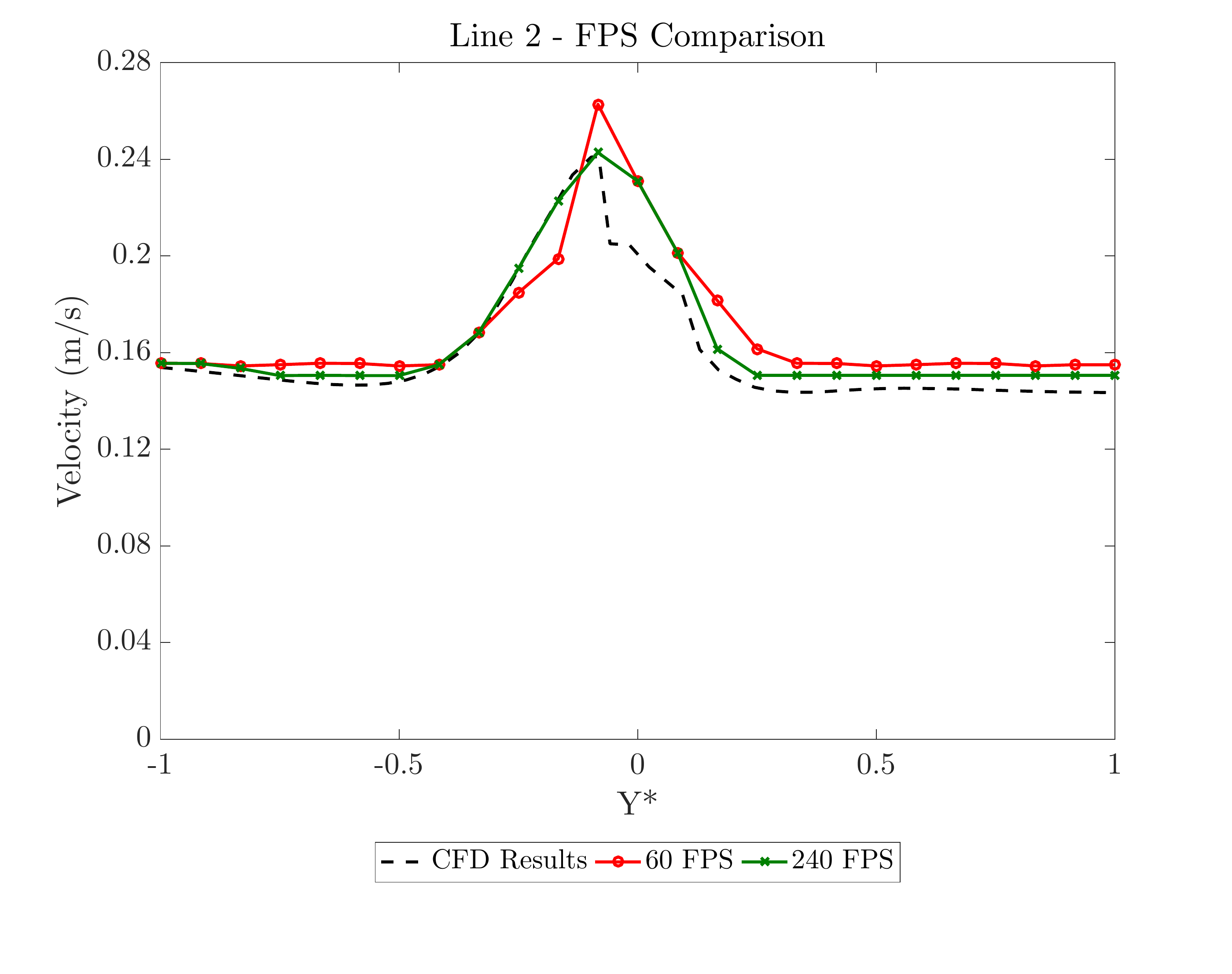}}
\end{minipage}\par\medskip
\centering
\subfloat[]{\label{f8c}\includegraphics[scale=.32]{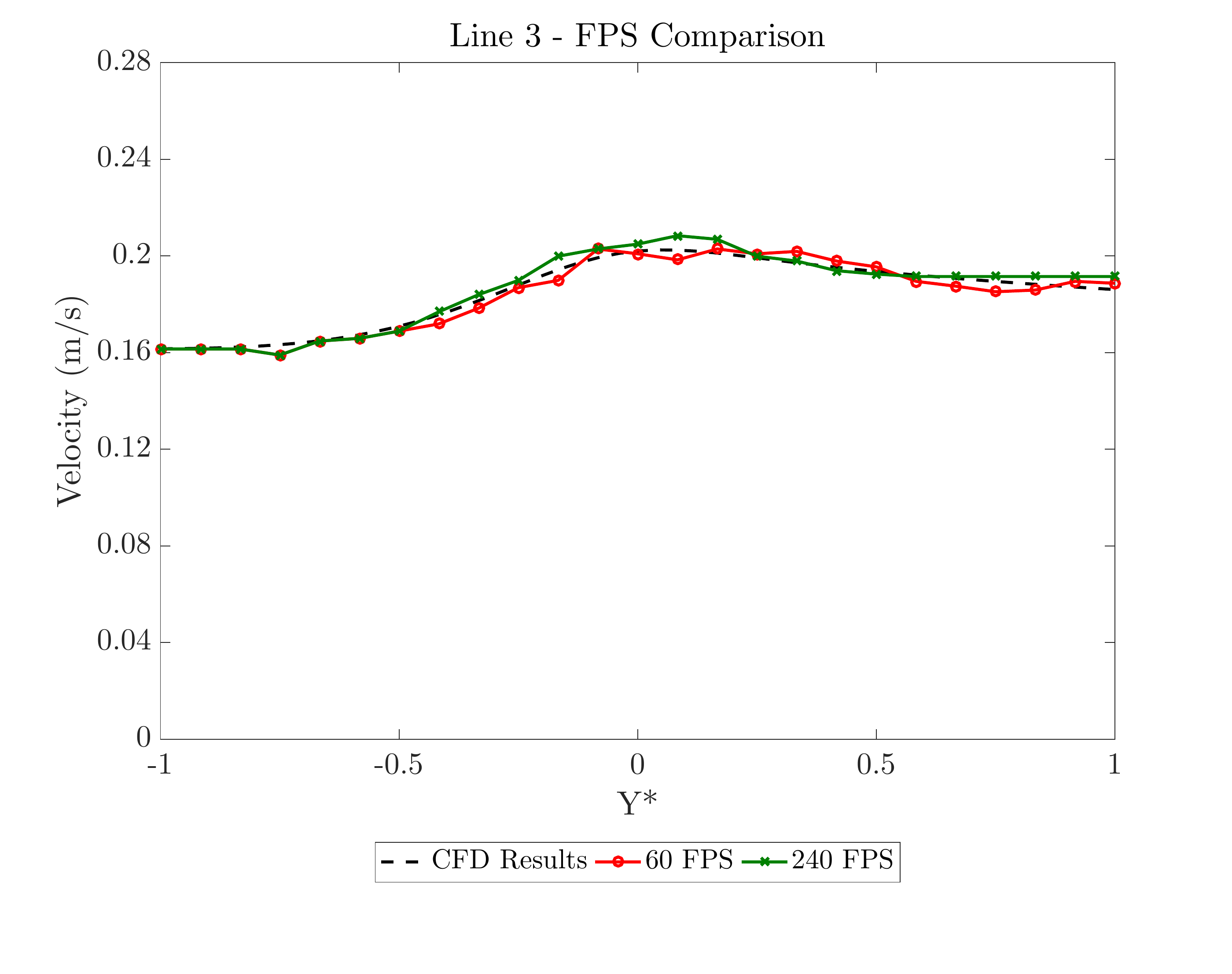}}
\caption{Values of velocities at various frame rates with change in Y-coordinate for M2}
\label{f8}
\end{figure}

The velocities obtained through experimental analysis were compared with CFD values obtained for the same line segments. The velocities were plotted against values of $Y^{*}$.
Figure \ref{f7} shows the observed velocities plotted against values of $Y^{*}$ for various values of frame rate for M1, while Figure \ref{f8} shows these values for M2.\\
\begin{table}[H]
\centering
\caption{\label{t2}Deviation at various values of FPS for M1 and M2}
\begin{tabular}{ |l|l|l|l| }
\hline
\textbf{Phone} & \textbf{Line} & \textbf{MAPE, 30 FPS} & \textbf{MAPE, 60 FPS} \\
\hline
M1 & 1 & 9.84\% & 7.18\%\\
\hline
M1 & 2 & 10.03\% & 6.32\%\\
\hline
M1 & 3 & 4.69\% & 1.09\%\\
\hline
M2 & 1 & 8.56\% & 2.32\%\\
\hline
M2 & 2 & 8.04\% & 3.54\%\\
\hline
M2 & 3 & 2.34\% & 1.06\%\\
\hline
\end{tabular}
\end{table}
\noindent It can be observed from the graphs that an increase in the frame rate leads to greater concordance with the CFD values for the study. Table \ref{t2} elaborates on the mean absolute percentage error (MAPE) of the obtained values from the benchmark values for the two phones. It is evident from the data presented that the measurements done using M2 at 240 FPS are most reliable, with the lowest values of MAPE for all 3 lines. The data also reflects the inference from the graphs, that an increase in FPS increases the concordance with CFD values. In their study on the use of CCD pixel binning in PIV, Akselli et al. \cite{s43} asserted that increasing the frame rate of the camera had a significant positive impact on the results, even if it was at the cost of spatial resolution. An increase in frame rate means that pictures are captured at closer intervals in time, which allows for more reliable measurement of the exact trajectory of the flow by the PIV software. This is particularly important for complex flows, where the velocity or direction of flow may change rapidly with change in geometry or physical conditions.\\

\subsection{Analysis of Exposure Compensation}
\begin{figure}[H]
\centering
\begin{minipage}{.5\linewidth}
\centering
\subfloat[]{\label{f9a}\includegraphics[scale=.32]{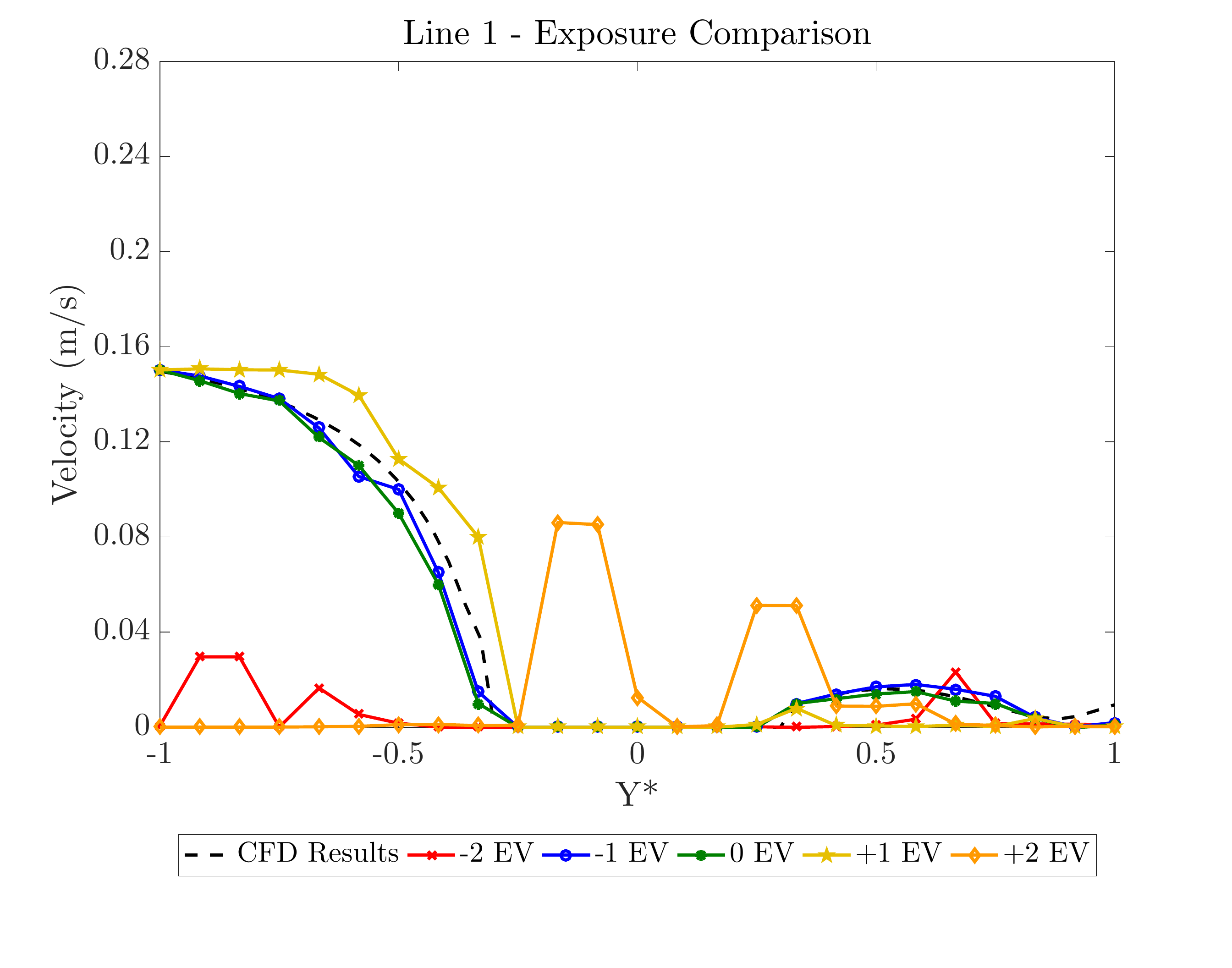}}
\end{minipage}%
\begin{minipage}{.5\linewidth}
\centering
\subfloat[]{\label{f9b}\includegraphics[scale=.32]{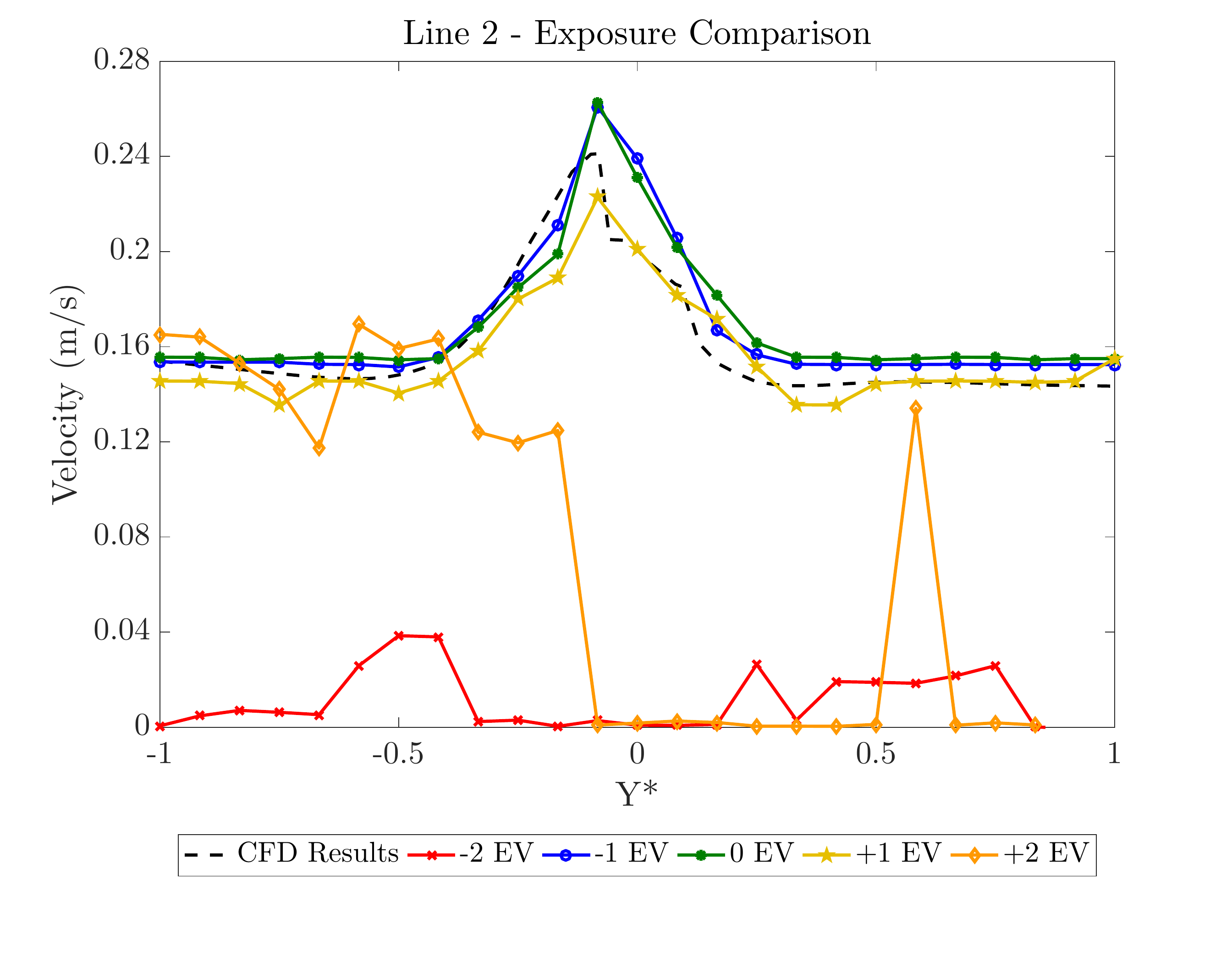}}
\end{minipage}\par\medskip
\centering
\subfloat[]{\label{f9c}\includegraphics[scale=.32]{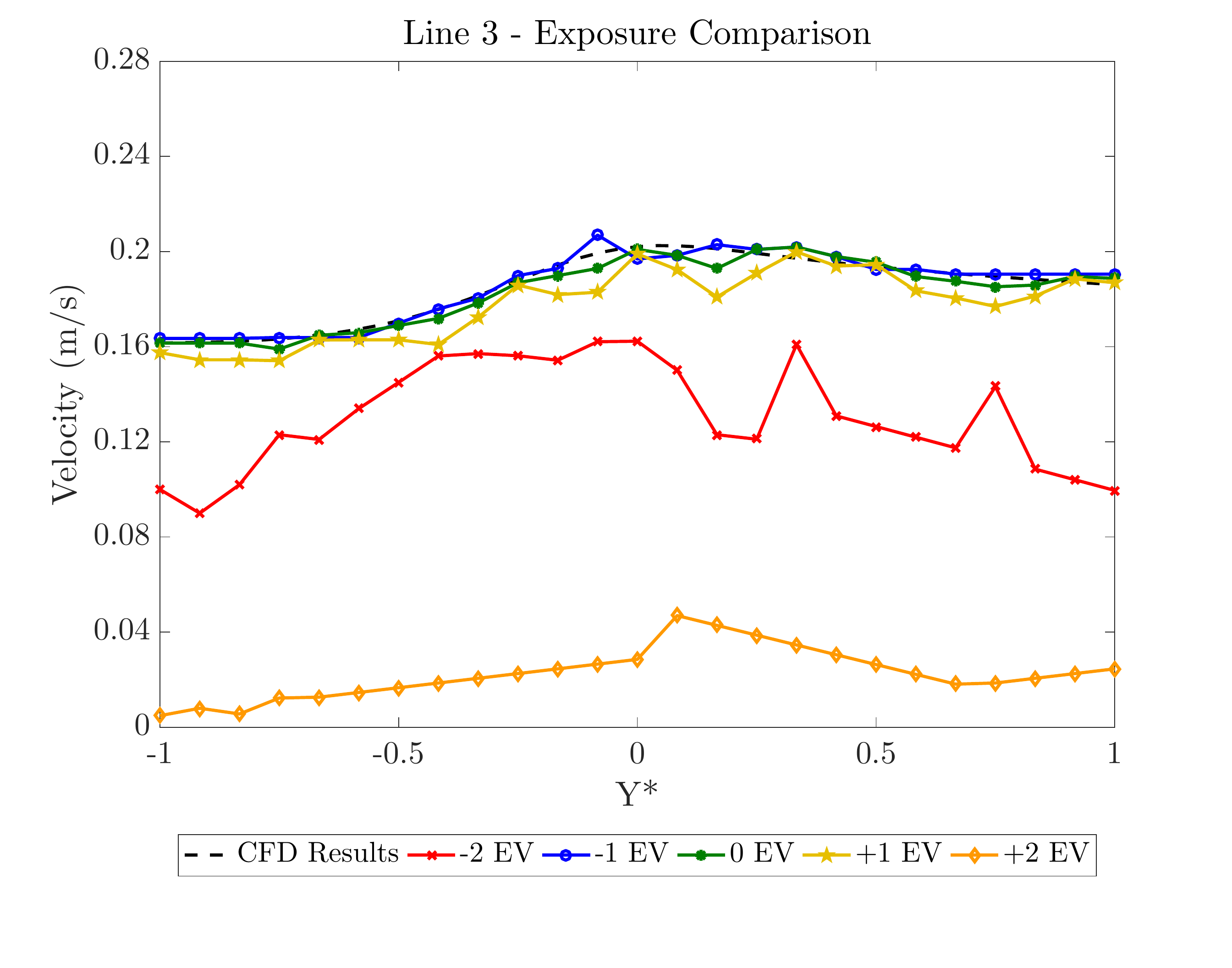}}
\caption{Values of velocities at various exposure compensations with change in Y-coordinate for M1}
\label{f9}
\end{figure}

\begin{figure}[H]
\centering
\begin{minipage}{.5\linewidth}
\centering
\subfloat[]{\label{f10a}\includegraphics[scale=.32]{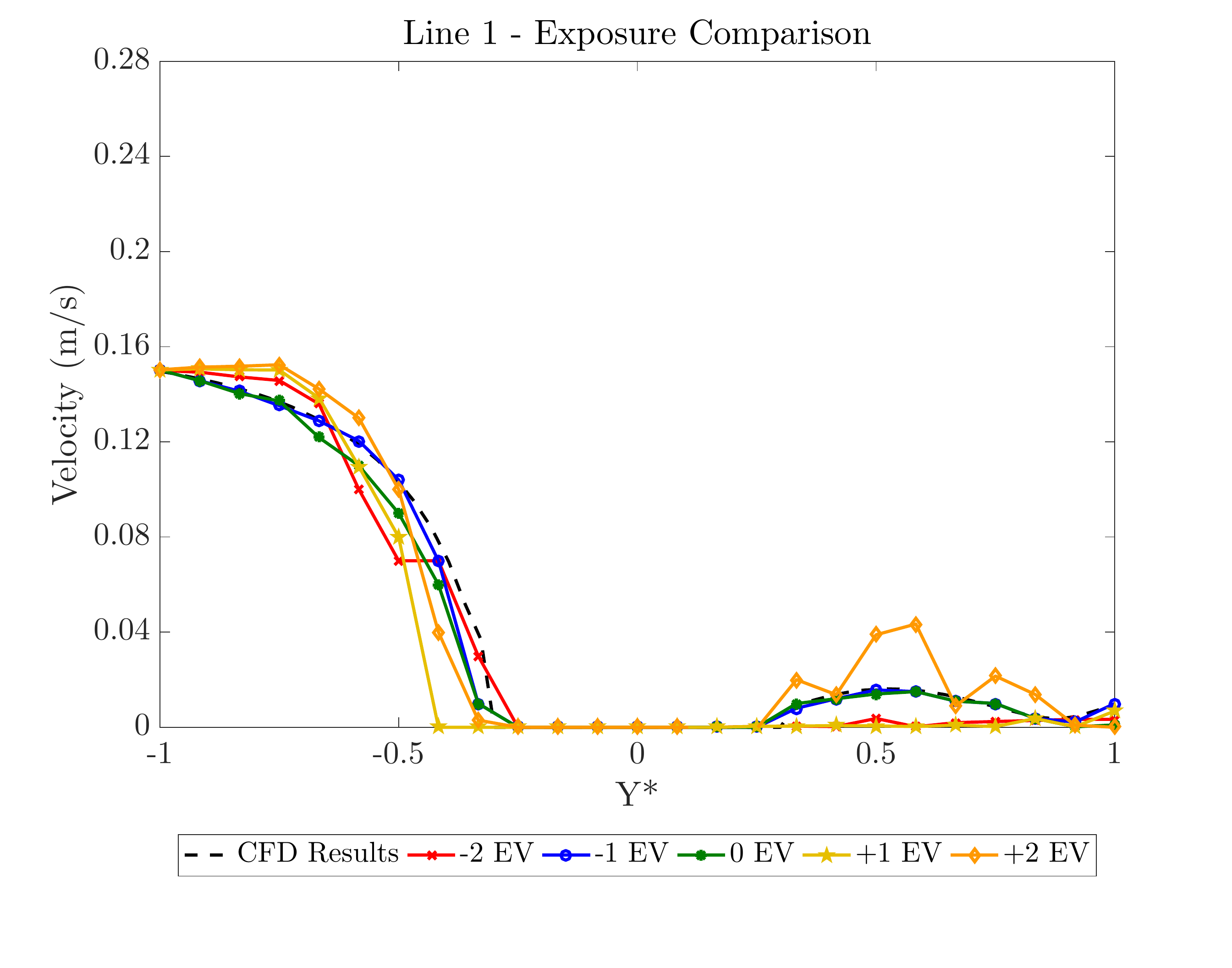}}
\end{minipage}%
\begin{minipage}{.5\linewidth}
\centering
\subfloat[]{\label{f10b}\includegraphics[scale=.32]{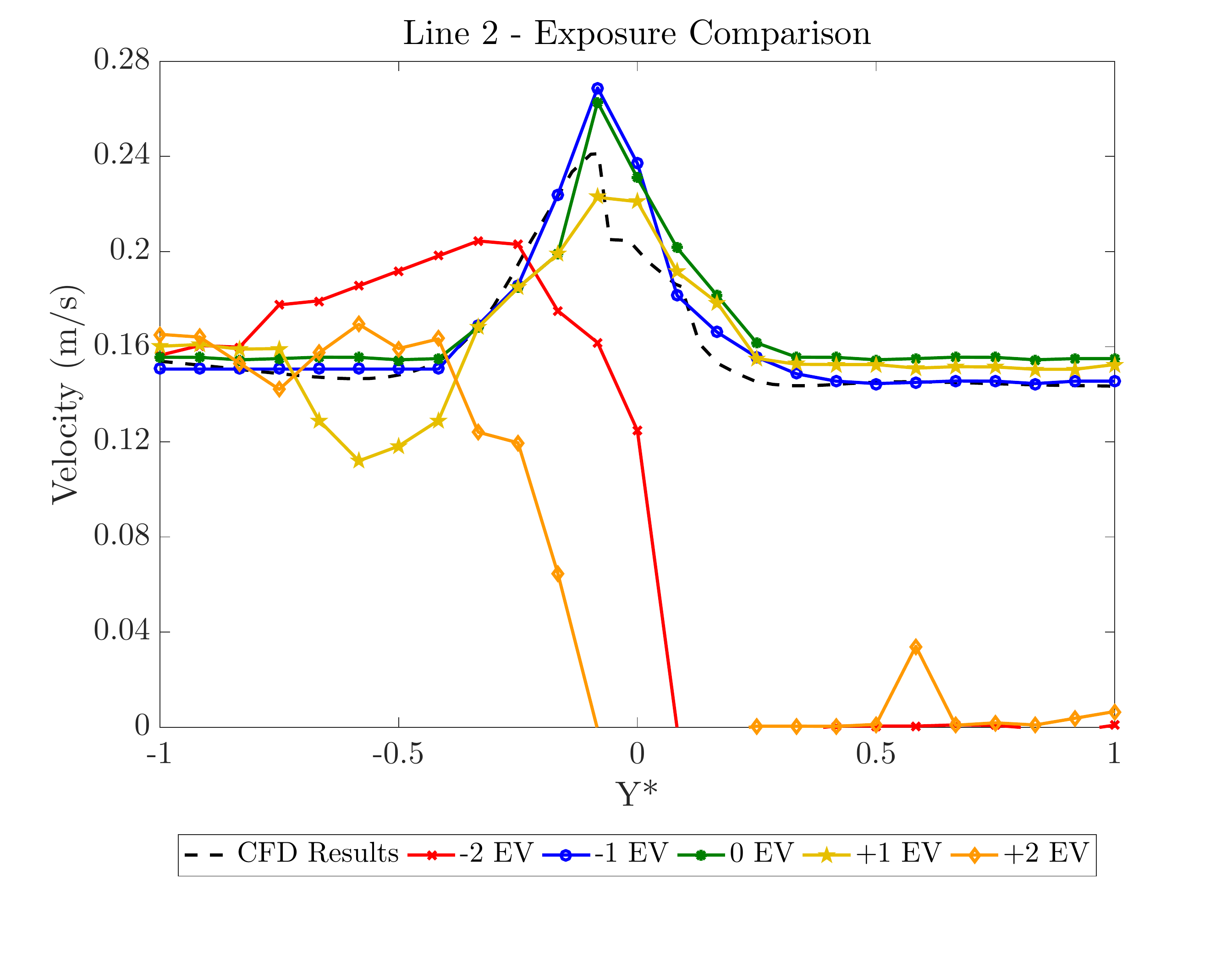}}
\end{minipage}\par\medskip
\centering
\subfloat[]{\label{f10c}\includegraphics[scale=.32]{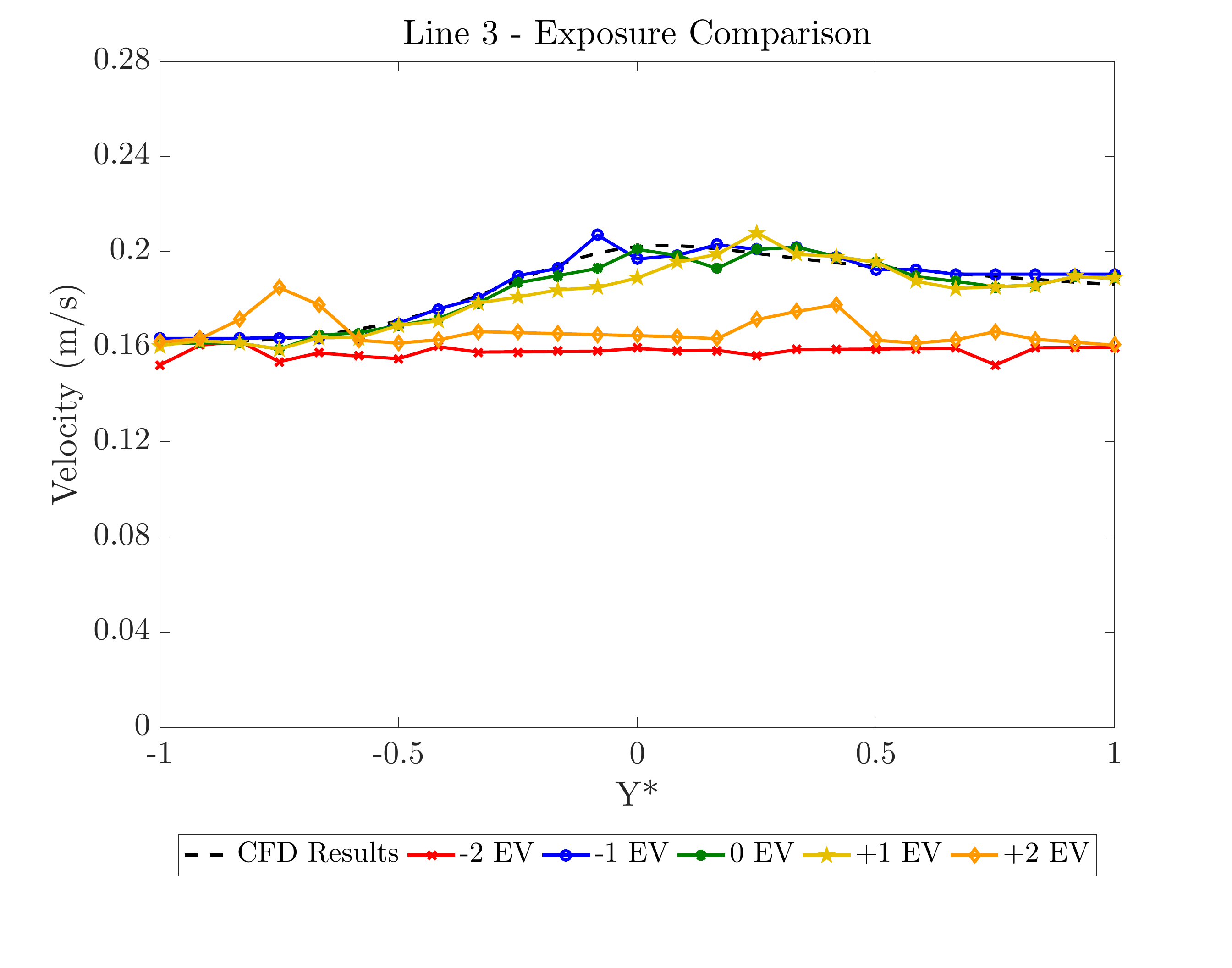}}
\caption{Values of velocities at various exposure compensations with change in Y-coordinate for M2}
\label{f10}
\end{figure}

Figure \ref{f9} shows the observed velocities plotted against values of $Y^{*}$ for various values of exposure compensation for M1, while Figure \ref{f10} contains these values for M2. The graphs indicate that the observations for exposure compensations of -2 and 2 are highly deviated, since they diverge considerably from the CFD values considered. The intermediate values of -1, 0 and 1 are observed to show comparatively greater agreement with these values. Table \ref{t3} depicts the MAPE of the obtained values from the benchmark values for the two phones.\\
\begin{table}[H]
\centering
\caption{\label{t3}Deviation at various values of Exposure Compensation for M1 and M2}
\begin{tabular}{ |l|l|l|l| }
\hline
\textbf{Phone} & \textbf{Line} & \textbf{MAPE, -1 EV} & \textbf{MAPE, 0 EV} \\
\hline
M1 & 1 & 6.08\% & 7.56\%\\
\hline
M1 & 2 & 5.68\% & 7.04\%\\
\hline
M1 & 3 & 1.07\% & 2.34\%\\
\hline
M2 & 1 & 2.47\% & 4.87\%\\
\hline
M2 & 2 & 3.29\% & 4.94\%\\
\hline
M2 & 3 & 0.99\% & 2.01\%\\
\hline
\end{tabular}
\end{table}
While it is evident from Figure \ref{f9} and Figure \ref{f10} that an exposure compensation of 1 EV gives significantly deviated results, the observations for -1 EV and 0 EV are observed to be relatively closer to CFD values. However, in each case studied, it was observed that an exposure compensation of -1 EV gave the greatest accuracy. Reducing the exposure time reduces the amount of light entering the camera. This also allows for greater autocorrelation, thus leading to more reliable results. However, reducing the exposure compensation beyond a point can lead to the amount of light not being enough to decipher even the seeding particles, which will lead to inaccurate results.
\subsection{Analysis of ISO levels}
\begin{figure}[H]
\centering
\begin{minipage}{.5\linewidth}
\centering
\subfloat[]{\label{f11a}\includegraphics[scale=.32]{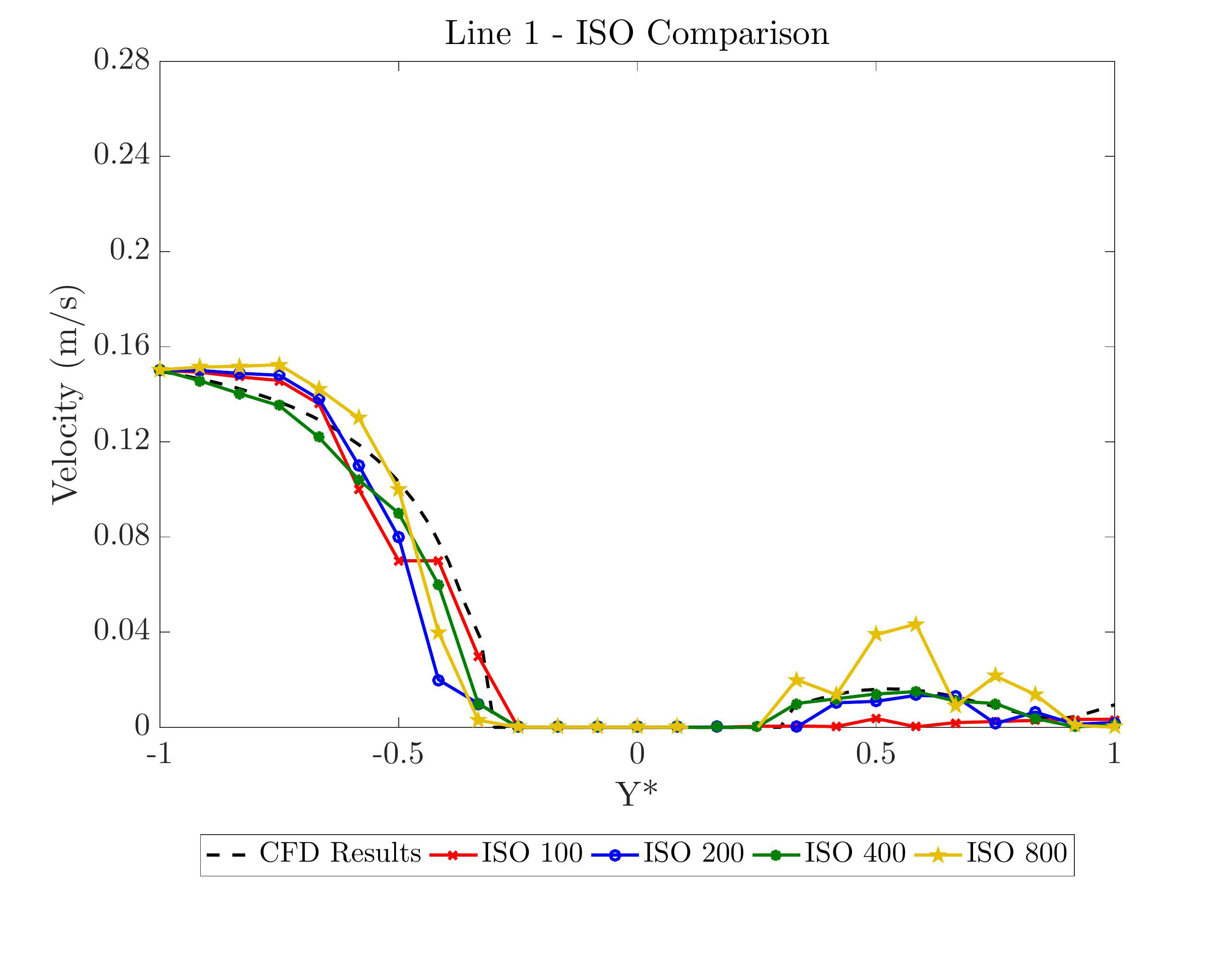}}
\end{minipage}%
\begin{minipage}{.5\linewidth}
\centering
\subfloat[]{\label{f11b}\includegraphics[scale=.32]{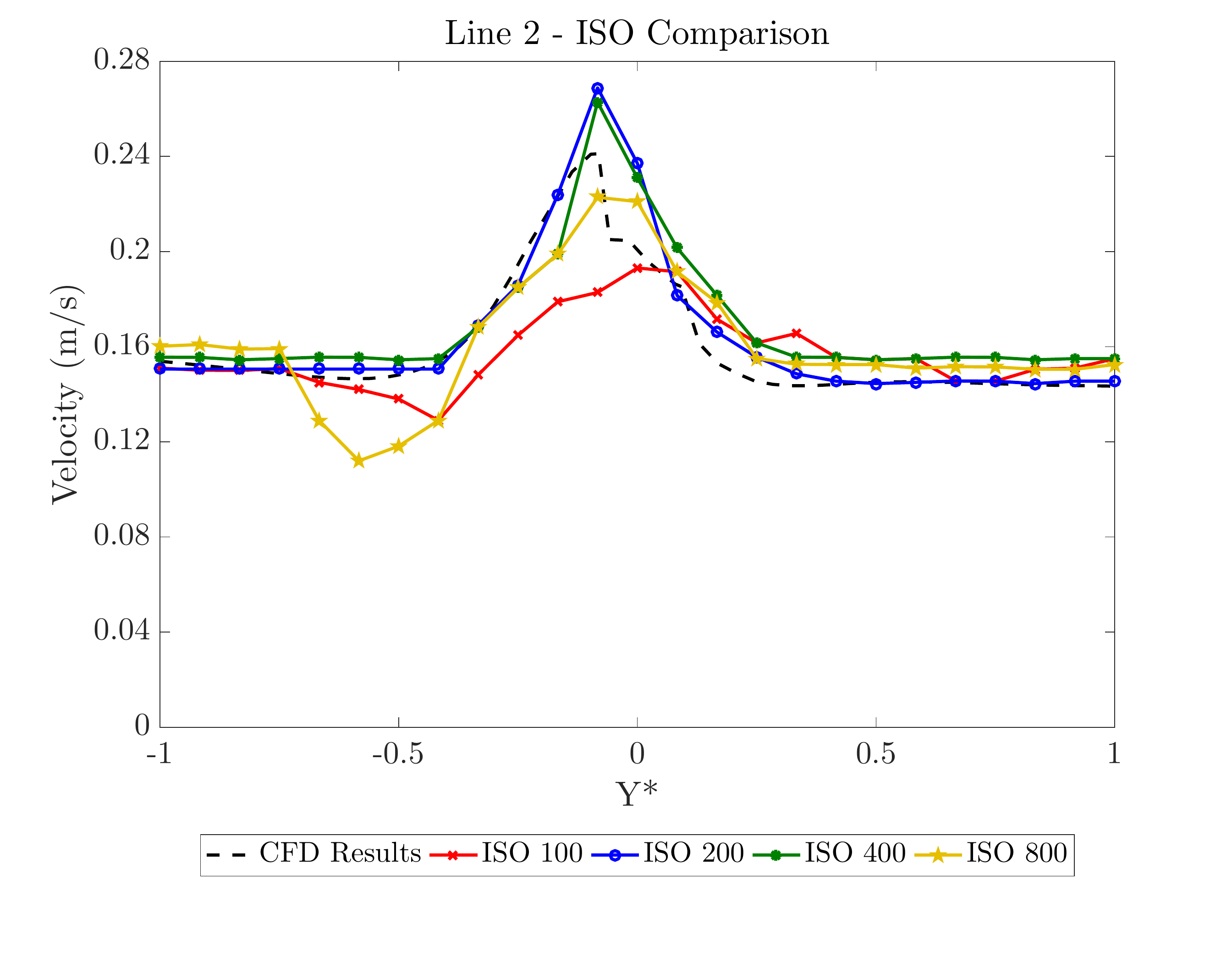}}
\end{minipage}\par\medskip
\centering
\subfloat[]{\label{f11c}\includegraphics[scale=.32]{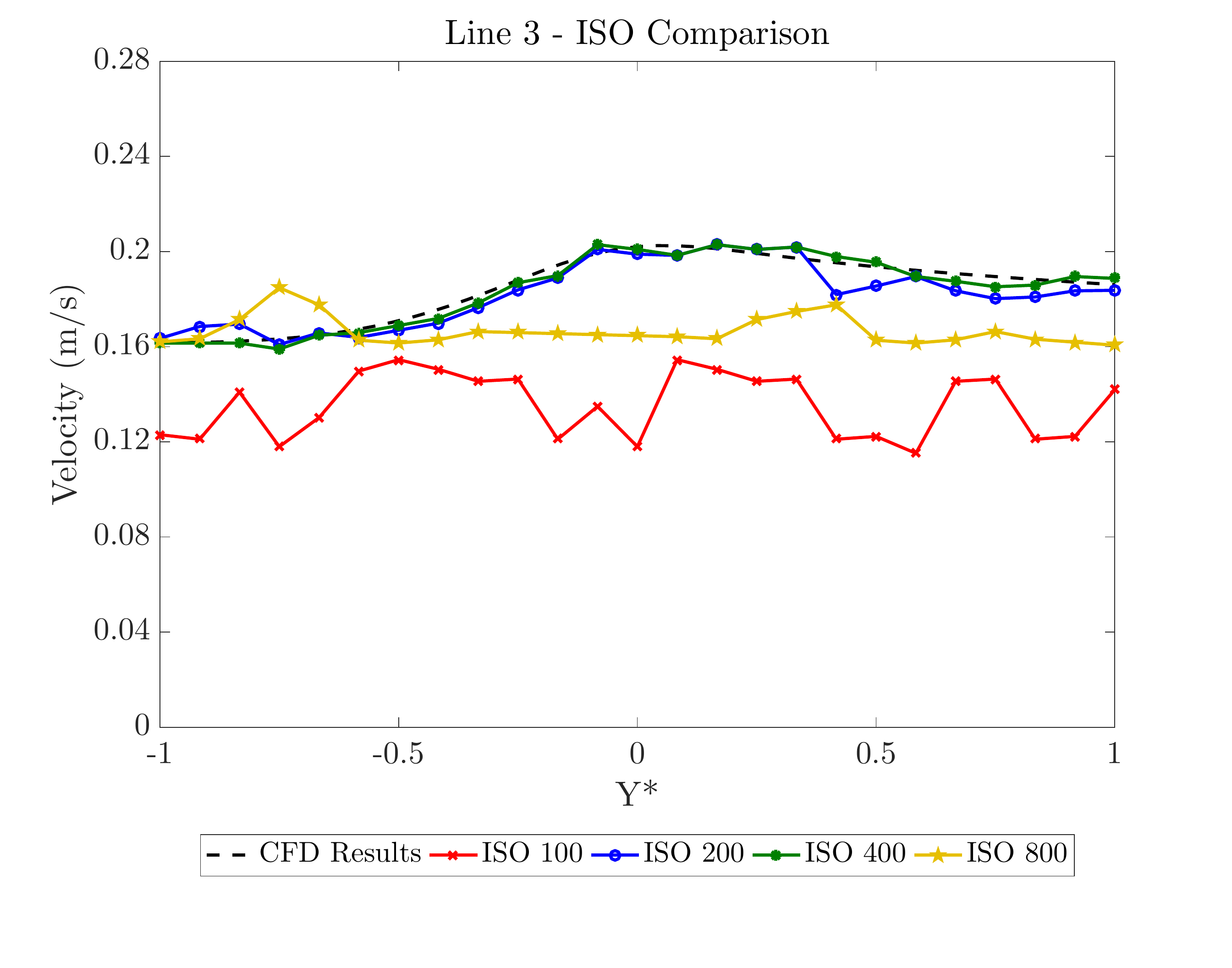}}
\caption{Values of velocities at various ISO levels with change in Y-coordinate for M1}
\label{f11}
\end{figure}

Figure \ref{f11} shows the PIV-measured velocities plotted against values of $Y^{*}$ for various values of ISO for M1. M2 does not have an option for adjustment of ISO levels. As in the case of exposure compensation, the extreme values of ISO are observed to be highly variant as compared to the CFD values for the analysis. Table \ref{t4} gives the MAPE of the obtained values from the benchmark values for the two phones. It is evident from the data obtained that an ISO of 400 gives the least deviated results, with a maximum deviation of about 6\%. ISO governs the brightness of an image, and higher values of ISO can lead to grainy images, which reduce the quality of the image and render seeding particles less decipherable. At the same time, lower values can lead to exceedingly dark images. The ISO value was reduced, and 400 was observed to be the most suitable.
\begin{table}[H]
\centering
\caption{\label{t4}Deviation at various values of ISO for M1}
\begin{tabular}{ |l|l|l|l| }
\hline
\textbf{Phone} & \textbf{Line} & \textbf{MAPE, 200 ISO} & \textbf{MAPE, 400 ISO} \\
\hline
M1 & 1 & 7.92\% & 5.23\%\\
\hline
M1 & 2 & 6.45\% & 5.15\%\\
\hline
M1 & 3 & 3.02\% & 0.99\%\\
\hline
\end{tabular}
\end{table}
%
\subsection{Analysis of Deviation}
\begin{figure}[H]
\centering
\includegraphics[width=0.4\textwidth]{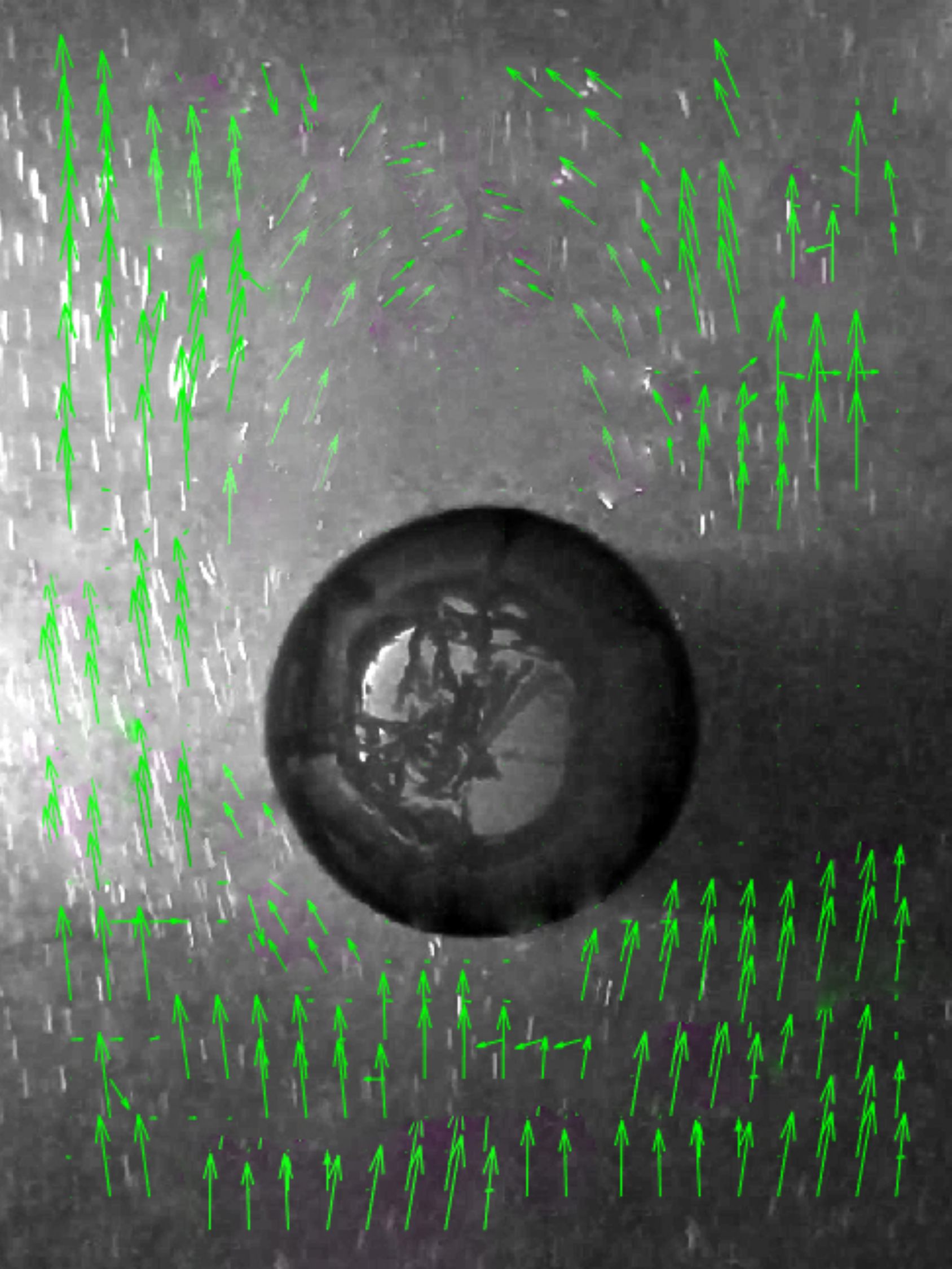}
\caption{Velocity vectors for analysis with M2 at 240 FPS }
\label{f12}
\end{figure}

The experimental observations were compared to the CFD results in each test conducted in the present study. The mean average deviation (MAD), standard deviation and mean absolute percentage errors(MAPE) were calculated, as shown in table \ref{t5}.\\
\begin{table}[H]
\centering
\caption{\label{t5}Deviation analysis of experimental observations.}
\begin{tabular}{ |p{0.05\linewidth}|p{0.06\linewidth}|p{0.06\linewidth}|p{0.25\linewidth}|p{0.1\linewidth}|p{0.25\linewidth}|p{0.1\linewidth}| }
\hline
\textbf{S.No.} & \textbf{Line} & \textbf{Phone} & \textbf{Characteristic} & \textbf{MAD (m/s)} & \textbf{Standard Deviation (m/s)} & \textbf{MAPE (\%)}\\
\hline
1 & Line 1 & M1 & Frame Rate (60 FPS) & 0.0041 & 0.0047 & 7.18\\
\hline
2 & Line 1 & M1 & Exposure Compensation (-1 EV) &0.0041& 0.0047& 7.08\\
\hline
3 & Line 1 & M1 & ISO (400)&  0.0031 & 0.0032  & 5.23\\
\hline
4 & Line 1 & M2 & Frame Rate (240 FPS) & 0.0013 & 0.0015 & 2.32\\
\hline
5 & Line 1 & M2 & Exposure Compensation (-1 EV) & 0.0013 & 0.0015 & 2.47\\
\hline
6 & Line 2 & M1 & Frame Rate (60 FPS) & 0.0108 & 0.0073 & 6.32\\
\hline
7 & Line 2 & M1 & Exposure Compensation (-1 EV) & 0.0108 & 0.0073 & 6.68\\
\hline
8 & Line 2 & M1 & ISO (400) & 0.0083 & 0.0074 & 5.15\\
\hline
9 & Line 2 & M2 & Frame Rate (240 FPS) & 0.0057 & 0.0053 & 3.54\\
\hline
10 & Line 2 & M2 & Exposure Compensation (-1 EV) & 0.0053 & 0.0080 & 3.29\\
\hline
11 & Line 3 & M1 & Frame Rate (60 FPS) & 0.0024 & 0.0019 & 1.09\\
\hline
12 & Line 3 & M1 & Exposure Compensation (-1 EV) & 0.0024 & 0.0019 & 1.07\\
\hline 
13 & Line 3 & M1 & ISO (400) & 0.0022 & 0.0018 & 0.99\\
\hline
14 & Line 3 & M2 & Frame Rate (240 FPS) & 0.0022 & 0.0019 & 1.06\\
\hline
15 & Line 3 & M2 & Exposure Compensation (-1 EV) & 0.0022 & 0.0018 & 0.99\\
\hline
\end{tabular}
\end{table}
It was observed that the results obtained for the M2 were more accurate than those for the M1 in every case. The percentage deviation for the M2 was found to vary between 1\% and 3.54\%, while that for the M1 varied between 1\% and 7.08\%. The values obtained for Line 3 showed considerable agreement with the CFD results, while those for Line 1 showed a comparatively higher deviation. The average deviation for the M2 was calculated as 2.30\% while that for the M1 was 4.51\%. Figure \ref{f12} shows the resolved velocity vectors for the best setting from M2. Vectors at two regions are not present because of the lack of illumination predominantly on the right side of test piece.

\section{Conclusion}
A PIV system for flow visualization and measurement was constructed in INR 17300 (under USD 250). The designed system was able to observe the flow field in the region under study. It was observed that an increase in the value of frame rate increases the agreement of measurement. Further, an exposure compensation of -1 EV and an ISO of 400 were associated with the greatest agreement of measurement in the current study. The deviation for the optimum characteristics ranges from 1\% to 3.5\% for iPhone X and 1\% to 7\% for OnePlus 5T.\\

The given study also creates scope for further improvement in the area of smartphone-based PIV. As the frame rate capability of smartphones increases, so will the capacity of the system to give better results. An increase in camera quality (in terms of video resolution) can also be associated with an ability to resolve smaller seeding particles, improving output.
\section{Acknowledgements}
The authors would like to thank Third phase of Technical Education Quality Improvement Programme (referred to as TEQIP-III) at Delhi Technological University (DTU) for providing grant to carry out procurement of items and fabrication of apparatus necessary for carrying out the investigation. We would also like to thank the members of Fluid Mechanics Group and staff members of Fluid Systems Laboratory, DTU for providing technical support during fabrication of the appratus and conduction of experiment.
\section{Ethical statement}
The authors declare that the contents of the paper are not published elsewhere or will not be submitted to any other place during the review process. We also have no conflict of interests. 
\section*{Appendix I}
The total cost of the apparatus in Indian Rupee is shown in Table \ref{t8}.
\begin{table}[H]
\centering
\caption{\label{t8}Total Cost of Apparatus}
\begin{tabular}{ |l|l|l| }
\hline
\textbf{S. No.} & \textbf{Component} & \textbf{Cost (INR)}\\
\hline
1 & Open Channel(Acrylic) & 6200\\
\hline
2 & $0.5$ HP Pump & 2700\\
\hline
3 & Reservoirs $(\times 2)$ & 2100\\
\hline
4 & Channel Frame & 2000\\
\hline
5 & Silicone and Epoxy & 1600\\
\hline
6 & 1-Phase Variac & 1600\\
\hline
7 & PVC Fitting and Pipes & 700\\
\hline
8 & Laser Diode & 200 \\
\hline
9 & Aluminium Electrodes & 200 \\
\hline
\space & \textbf{Total} & 17300\\
\hline
\end{tabular}
\end{table}


\begin{thebibliography}{10}

\bibitem{s1}
M.~Raffel, C.~E. Willert, F.~Scarano, C.~J. K{\"a}hler, S.~T. Wereley, and
  J.~Kompenhans, {\em Particle image velocimetry: a practical guide}.
\newblock Springer, 2018.

\bibitem{s2}
A.~K. Prasad, ``Particle image velocimetry,'' {\em Current Science-Bangalore-},
  vol.~79, no.~1, pp.~51--60, 2000.

\bibitem{s3}
F.~Scarano, ``Tomographic piv: principles and practice,'' {\em Measurement
  Science and Technology}, vol.~24, no.~1, p.~012001, 2012.

\bibitem{s4}
K.~Lasinger, C.~Vogel, T.~Pock, and K.~Schindler, ``Variational 3d-piv with
  sparse descriptors,'' {\em Measurement Science and Technology}, vol.~29,
  no.~6, p.~064010, 2018.

\bibitem{s5}
W.~G. Ryerson and K.~Schwenk, ``A simple, inexpensive system for digital
  particle image velocimetry (dpiv) in biomechanics,'' {\em Journal of
  Experimental Zoology Part A: Ecological Genetics and Physiology}, vol.~317,
  no.~2, pp.~127--140, 2012.

\bibitem{s6}
C.~Gray, ``The development of particle image velocimetry for water wave
  studies.,'' 1989.

\bibitem{s7}
R.~J. Adrian, ``Twenty years of particle image velocimetry,'' {\em Experiments
  in fluids}, vol.~39, no.~2, pp.~159--169, 2005.

\bibitem{s8}
S.~Discetti and R.~J. Adrian, ``High accuracy measurement of magnification for
  monocular piv,'' {\em Measurement Science and Technology}, vol.~23, no.~11,
  p.~117001, 2012.

\bibitem{s9}
S.~J. Beresh, J.~F. Henfling, R.~W. Spillers, and S.~M. Spitzer,
  ``‘postage-stamp piv’: small velocity fields at 400 khz for turbulence
  spectra measurements,'' {\em Measurement Science and Technology}, vol.~29,
  no.~3, p.~034011, 2018.

\bibitem{s10}
S.~Scharnowski, A.~Sciacchitano, and C.~Kaehler, ``On the universality of keane
  \& adrian's valid detection probability in piv,'' {\em Measurement Science
  and Technology}, 2019.

\bibitem{s11}
B.~P. Ring, D.~K. Atkinson, A.~W. Henderson, and E.~C. Lemley, ``Development of
  a low cost particle image velocimetry system for fluids engineering research
  and education,'' in {\em ASME 2013 Fluids Engineering Division Summer
  Meeting}, American Society of Mechanical Engineers Digital Collection, 2013.

\bibitem{s12}
D.~Mei, J.~Ding, S.~Shi, T.~New, and J.~Soria, ``High resolution volumetric
  dual-camera light-field piv,'' {\em Experiments in Fluids}, vol.~60, no.~8,
  p.~132, 2019.

\bibitem{s13}
E.~Longmire and A.~Alahyari, ``Piv measurements in simulated microbursts,'' in
  {\em Proceedings 7th International Symposium on Applications of Laser
  Techniques to Fluid Mechanics, Lisbon}, 1994.

\bibitem{s14}
Z.-C. Liu, C.~Landreth, R.~Adrian, and T.~Hanratty, ``High resolution
  measurement of turbulent structure in a channel with particle image
  velocimetry,'' {\em Experiments in fluids}, vol.~10, no.~6, pp.~301--312,
  1991.

\bibitem{s15}
B.~Khoo, T.~Chew, P.~Heng, and H.~Kong, ``Turbulence characterisation of a
  confined jet using piv,'' {\em Experiments in fluids}, vol.~13, no.~5,
  pp.~350--356, 1992.

\bibitem{s16}
M.~Bown, J.~MacInnes, R.~Allen, and W.~Zimmerman, ``Three-dimensional,
  three-component velocity measurements using stereoscopic micro-piv and ptv,''
  {\em Measurement Science and Technology}, vol.~17, no.~8, p.~2175, 2006.

\bibitem{s17}
L.~J. Graham and J.~Soria, ``A study of an inclined cylinder wake using digital
  particle image velocimetry,'' in {\em Proceedings 7th International Symposium
  on Applications of Laser Techniques to Fluid Mechanics, Lisbon}, 1994.

\bibitem{s18}
A.~Melling, ``Tracer particles and seeding for particle image velocimetry,''
  {\em Measurement Science and Technology}, vol.~8, no.~12, p.~1406, 1997.

\bibitem{s19}
T.~Van~Overbr{\"u}ggen, M.~Klaas, J.~Soria, and W.~Schr{\"o}der, ``Experimental
  analysis of particle sizes for piv measurements,'' {\em Measurement Science
  and Technology}, vol.~27, no.~9, p.~094009, 2016.

\bibitem{s20}
J.~Dieter, R.~Bremeyer, F.~Hering, and B.~J{\"a}hne, ``Flow measurements close
  to the free air/sea interface,'' in {\em Proceedings 7th International
  Symposium on Applications of Laser Techniques to Fluid Mechanics, Lisbon},
  1994.

\bibitem{s21}
F.~Pedocchi, J.~E. Martin, and M.~H. Garc{\'\i}a, ``Inexpensive fluorescent
  particles for large-scale experiments using particle image velocimetry,''
  {\em Experiments in Fluids}, vol.~45, no.~1, pp.~183--186, 2008.

\bibitem{s22}
M.~Walkden, G.~M{\"u}ller, T.~Bruce, {\em et~al.}, ``Low-cost particle image
  velocimetry: system and application,'' in {\em The Eighth International
  Offshore and Polar Engineering Conference}, International Society of Offshore
  and Polar Engineers, 1998.

\bibitem{s23}
D.~P. Hart, ``Sparse array image correlation,'' in {\em Developments in Laser
  Techniques and Fluid Mechanics}, pp.~53--74, Springer, 1997.

\bibitem{s24}
Y.~Hassan, R.~Martinez, O.~Philip, and W.~Schmidl, ``Flow measurement of a
  two-phase fluid around a cylinder in a channel using particle image
  velocimetry,'' {\em Transactions of the American Nuclear Society}, vol.~71,
  no.~CONF-941102-, 1994.

\bibitem{s25}
M.~Oishi, H.~Kinoshita, T.~Fujii, and M.~Oshima, ``Phase-locked confocal
  micro-piv measurement for 3d flow structure of transient droplet formation
  mechanism in t-shaped microjunction,'' {\em Measurement Science and
  Technology}, vol.~29, no.~11, p.~115204, 2018.

\bibitem{s26}
I.~Grant, G.~Smith, D.~Infield, X.~Wang, Y.~Zhao, and S.~Fu, ``Measurements of
  the flow around wind turbine rotors by particle image velocimetry,'' in {\em
  Proc. 7th Int. Symp. on Applications of Laser Techniques to Fluid Mechanics,
  Lisbon}, 1994.

\bibitem{s27}
I.~Grant and X.~Wang, ``Directionally-unambiguous, digital particle image
  velocimetry studies using a image intensifier camera,'' {\em Experiments in
  fluids}, vol.~18, no.~5, pp.~358--362, 1995.

\bibitem{s28}
C.~Cierpka, R.~Hain, and N.~A. Buchmann, ``Flow visualization by mobile phone
  cameras,'' {\em Experiments in Fluids}, vol.~57, no.~6, p.~108, 2016.

\bibitem{s29}
C.~Willert, B.~Stasicki, J.~Klinner, and S.~Moessner, ``Pulsed operation of
  high-power light emitting diodes for imaging flow velocimetry,'' {\em
  Measurement Science and Technology}, vol.~21, no.~7, p.~075402, 2010.

\bibitem{s30}
N.~A. Buchmann, C.~E. Willert, and J.~Soria, ``Pulsed, high-power led
  illumination for tomographic particle image velocimetry,'' {\em Experiments
  in fluids}, vol.~53, no.~5, pp.~1545--1560, 2012.

\bibitem{s31}
A.~A. Aguirre-Pablo, M.~K. Alarfaj, E.~Q. Li, J.~F. Hern{\'a}ndez-S{\'a}nchez,
  and S.~T. Thoroddsen, ``Tomographic particle image velocimetry using
  smartphones and colored shadows,'' {\em Scientific reports}, vol.~7, no.~1,
  p.~3714, 2017.

\bibitem{s32}
M.~Mommert, D.~Schiepel, D.~Schmeling, and C.~Wagner, ``A flow-intrinsic
  trigger for capturing reconfigurations in buoyancy-driven flows in automated
  piv,'' {\em Measurement Science and Technology}, vol.~30, no.~4, p.~045301,
  2019.

\bibitem{s33}
M.~B.~P. Ring and E.~C. Lemley, ``Design and implementation of a low cost
  particle image velocimetry sys-tem for undergraduate research and
  education,'' {\em age}, vol.~24, p.~1, 2014.

\bibitem{s42}
P.~J. Pritchard and J.~W. Mitchell, {\em Fox and McDonald's Introduction to
  Fluid Mechanics, Binder Ready Version}.
\newblock John Wiley \& Sons, 2016.

\bibitem{s34}
M.~S. Kirkg{\"o}z and M.~Ardi{\c{c}}lio{\u{g}}lu, ``Velocity profiles of
  developing and developed open channel flow,'' {\em Journal of Hydraulic
  Engineering}, vol.~123, no.~12, pp.~1099--1105, 1997.

\bibitem{s35}
H.~Bonakdari, F.~Larrarte, L.~Lassabatere, and C.~Joannis, ``Turbulent velocity
  profile in fully-developed open channel flows,'' {\em Environmental Fluid
  Mechanics}, vol.~8, no.~1, pp.~1--17, 2008.

\bibitem{s37}
Z.~J. Taylor, R.~Gurka, G.~A. Kopp, and A.~Liberzon, ``Long-duration
  time-resolved piv to study unsteady aerodynamics,'' {\em IEEE Transactions on
  Instrumentation and Measurement}, vol.~59, no.~12, pp.~3262--3269, 2010.

\bibitem{s38}
F.~Durst, S.~Ray, B.~{\"U}nsal, and O.~Bayoumi, ``The development lengths of
  laminar pipe and channel flows,'' {\em Journal of fluids engineering},
  vol.~127, no.~6, pp.~1154--1160, 2005.

\bibitem{s39}
S.-Y. Hsu, ``Effects of water flow rate, salt concentration and water
  temperature on efficiency of an electrolyzed oxidizing water generator,''
  {\em Journal of Food Engineering}, vol.~60, no.~4, pp.~469--473, 2003.

\bibitem{s40}
``Dvdvideosoft,'' 2018.

\bibitem{gti}
S.~Kumar, P.~Gupta, and R.~K. Singh, ``{Metaheuristic Optimization of
  Dual-Element Vertical Axis Wind Turbine Using Genetic Algorithm},'' vol.~2:
  Renewable Energy: Solar and Wind of {\em ASME Gas Turbine India Conference},
  12 2019.

\bibitem{gaslat}
S.~Kumar, P.~Gupta, and R.~K. Singh, ``{A Natural Evolution Based Numerical
  Optimisation Framework to Develop and Enhance Airfoil-Slat Arrangement},''
  vol.~7: Fluids Engineering of {\em ASME International Mechanical Engineering
  Congress and Exposition}, 11 2019.

\bibitem{mru}
A.~Saxena, E.~Ng, M.~Mathur, C.~Manchanda, and N.~A. Jajal, ``Effect of carotid
  artery stenosis on neck skin tissue heat transfer,'' {\em International
  Journal of Thermal Sciences}, vol.~145, p.~106010, 2019.

\bibitem{s41}
H.~Schlichting and K.~Gersten, {\em Boundary-layer theory}.
\newblock Springer, 2016.

\bibitem{s43}
B.~Akselli, A.~Kholmatov, and H.~Nasibov, ``The use of ccd pixel binning in piv
  measurements,'' in {\em 2009 International Symposium on Optomechatronic
  Technologies}, pp.~223--228, IEEE, 2009.

\end{thebibliography}
\end{document}